\documentclass[12pt]{article} 
\usepackage{amssymb}
\usepackage[dvips]{graphicx}
\newcommand{\NC}{\newcommand}

\newcommand{\newsubsection}[1]{
\vspace{7mm}
\pagebreak[3]
\addtocounter{subsection}{1}
\addcontentsline{toc}{subsection}{\protect
\numberline{\arabic{section}.\arabic{subsection}}{#1}}
\noindent{\it \thesubsection.  #1}                 
\nopagebreak
\vspace{2mm}
\nopagebreak}

\newcommand{\newsection}[1]{
\vspace{15mm}
\pagebreak[3]
\addtocounter{section}{1}
\setcounter{equation}{0}
\setcounter{subsection}{0}
\addcontentsline{toc}{section}{\protect
\numberline{\arabic{section}}{{\rm #1}}}
\noindent{\bf \thesection.  #1}                 
\nopagebreak
\vspace{4mm}
\nopagebreak}

\newcommand{\appsection}[1]{
\vspace{15mm}
\pagebreak[3]
\addtocounter{section}{1}
\setcounter{equation}{0}
\setcounter{subsection}{0}
\addcontentsline{toc}{section}{\protect
\numberline{\Alph{section}}{{\rm #1}}}
\noindent{\bf Appendix \thesection.  #1}                 
\nopagebreak
\vspace{4mm}
\nopagebreak}

\renewcommand{\theequation}{\thesection.\arabic{equation}}
\newlength{\extraspace}
\newlength{\extraspaces}
\newcounter{dummy}
\newcommand{\be}{\begin{equation}
\addtolength{\abovedisplayskip}{\extraspaces}
\addtolength{\belowdisplayskip}{\extraspaces}
\addtolength{\abovedisplayshortskip}{\extraspace}
\addtolength{\belowdisplayshortskip}{\extraspace}}
\newcommand{\ee}{\end{equation}}
\newcommand{\ba}{\begin{eqnarray}
\addtolength{\abovedisplayskip}{\extraspaces}
\addtolength{\belowdisplayskip}{\extraspaces}
\addtolength{\abovedisplayshortskip}{\extraspace}
\addtolength{\belowdisplayshortskip}{\extraspace}}
\newcommand{\ea}{\end{eqnarray}}
\newcommand{\nonu}{\nonumber \\[2mm]}
\newcommand{\is}{& \!\! = \!\! &}
\newcommand{\ban}{\begin{eqnarray*}
\addtolength{\abovedisplayskip}{\extraspaces}
\addtolength{\belowdisplayskip}{\extraspaces}
\addtolength{\abovedisplayshortskip}{\extraspace}
\addtolength{\belowdisplayshortskip}{\extraspace}}
\newcommand{\ean}{\end{eqnarray*}}
\newcommand{\baa}{                         
\addtocounter{equation}{1}
\setcounter{dummy}{\value{equation}}
\setcounter{equation}{0}
\renewcommand{\theequation}{\thesection.\arabic{dummy}\alph{equation}}
\begin{eqnarray}
\addtolength{\abovedisplayskip}{\extraspaces}
\addtolength{\belowdisplayskip}{\extraspaces}
\addtolength{\abovedisplayshortskip}{\extraspace}
\addtolength{\belowdisplayshortskip}{\extraspace}}
\newcommand{\eaa}{                                       
\end{eqnarray}
\setcounter{equation}{\value{dummy}}
\renewcommand{\theequation}{\thesection.\arabic{equation}}}

\NC{\fig}[3]{\protect\raisebox{#3 pt}{\protect\includegraphics[scale=#2]{#1.eps}}}

\renewcommand{\d}{{\partial}}
\newcommand{\fr}[1]{{\textstyle{1\over #1}}}

\newcommand{\Z}{{\mathbb Z}}
\newcommand{\RR}{{\mathbb R}}

\newcommand{\cA}{{\cal A }}

\newcommand{\Tr}{{\rm Tr}\,}

\newcommand{\xbar}{{\overline{x}}}

\newcommand{\Cbar}{{\overline{C}}}

\newcommand{\Ebar}{{\overline{E}}}
\newcommand{\Fbar}{{\overline{F}}}

\newcommand{\Lbar}{{\overline{L}}}
\newcommand{\Mbar}{{\overline{M}}}

\newcommand{\Rbar}{{\overline{R}}}

\newcommand{\Wbar}{{\overline{W}}}

\newcommand{\Psibar}{{\overline{\Psi}}}

\newcommand{\cD}{{\cal D}}

\newcommand{\cN}{{\cal N}}

\newcommand{\Diff}{{\rm Diff}}

\def\a{\alpha} 
\def\b{\beta} 
\def\g{\gamma} 
\def\G{\Gamma}
\def\e{\varepsilon}
\def\h{\eta}

\def\l{\lambda} 
\def\L{\Lambda} 
\def\m{\mu}
\def\n{\nu}
 
\def\s{\sigma} 
\def\t{\tau}
\def\f{\phi} 
\def\F{\Phi} 

\def\W{\Omega}

\def\<{\langle}
\def\>{\rangle}

\newfont{\gothic}{eufm10 scaled\magstep1}

\NC{\Seff}{S_{\rm eff}}
\NC{\Scl}{S_{\rm cl}}
\NC{\Sa}{S_{AdS}}
\NC{\Am}{A_{\rm min}}
\NC{\St}{S_{\rm II}}
\NC{\rf}[1]{(\ref{#1})}
\NC{\dis}{\displaystyle}
\NC{\MM}{\Mbar^\mu_{abcd}}
\NC{\LL}{\Lbar^\mu_{abc}}
\NC{\Fm}{\Fbar^\mu}
\NC{\Em}{\Ebar^\mu}
\NC{\Lm}{\Lbar^\mu}
\NC{\Mm}{\Mbar^\mu}
\NC{\Wm}[1]{\overline{\W}^\mu_{#1}}
\NC{\Wn}[1]{\overline{\W}^\nu_{#1}}
\NC{\Pm}{\Psibar^\mu}
\renewcommand{\Wbar}{\overline{\W}}
\NC{\D}{\nabla}
\NC{\gd}{\delta}
\NC{\dint}{\int\!\!\!\!\!\int}
\NC{\Ks}{K_{\s\s'}}
\NC{\Gf}[1]{G_{\rm fin}^{#1}}
\NC{\Gfin}{G_{\rm fin}}
\NC{\R}[1]{\Rbar_{#1}}
\renewcommand{\ll}{\log\L}
\newcommand{\sign}{{\mathop{\rm sign}}}
\NC{\lr}{\leftrightarrow}
\NC{\Lhat}{\widehat{L}}
\NC{\Cdbar}{\overline{\Cbar}}
\NC{\sump}{\mathop{{\sum}'}}
\renewcommand{\Diff}{Dif\hspace{-1pt}f}

\begin{document}

\addtolength{\baselineskip}{.5mm}
\thispagestyle{empty}
\begin{flushright}
PUPT-2036
\end{flushright}

\vspace{8mm}

\begin{center}
{\Large \sc 
Wilson Loops, D-branes, and \\[4mm] Reparametrization Path-integrals}
\\[25mm] {Vyacheslav S. Rychkov\footnote{\tt rytchkov@math.princeton.edu}}\\[4mm]
{\it Department of  Mathematics}\\[2mm]
{\cal \&}\\[2mm]
{\it Joseph Henry Laboratories\\ 
Princeton University\\ Princeton, NJ 08540}\\[2cm] 
{\sc Abstract}
\end{center}


\noindent We study path-integrals over reparametrizations of the 
world-sheet boundary. Such integrals arise when string propagates 
between fixed space-time contours. In gauge/string duality they 
are needed to describe gauge theory Wilson loops.
We show that (1) in AdS/CFT, the integral is well defined and 
gives a finite 1-loop correction to the Wilson
loop; (2) in critical string theory, the integral is UV divergent, and
fixed contour amplitudes are off shell. 
In the second case, we show that the divergences can be removed 
by renormalizing the contour. We calculate the 2-loop contour
$\b$-function and explain how it is related to the D0-brane effective action.
We also apply this method to compute the first $\a'$
correction to the effective action of higher dimensional branes.

\vspace{5mm}

\noindent April 2002
\vfill

\newpage

\newsection{Introduction}

Gauge fields are believed to have a dual description in terms of
strings, representing their flux lines. This belief has been
considerably strengthened in recent years, although a systematic 
theory is still missing. The most important problem is to move beyond
the supergravity approximation on string theory side of the
duality. Sooner or later it has to be done, if our final goal is to
describe physically interesting asymptotically free theories.

The gauge theory Wilson loop $W[C]$ has played an important role in discussions of
gauge/string duality, both in the early period \cite{Wilson:1974sk,Polyakov:1980ca} and in the recent work \cite{Polyakov:1998tj,Polyakov:1998ju}.
In the dual picture it has a very suggestive description as a
path-integral over string world-sheets with a fixed boundary contour $C$.
In the supergravity limit, this integral is dominated by the
corresponding minimal surface. This paper grew out of attempts to
understand quantum corrections to this supergravity result.

 A starting point of our discussion is an old observation
\cite{Polyakov:1987ez,Cohen:1986sm} that an amplitude describing string
propagating between fixed space-time contours reduces in conformal
gauge to a path-integral over boundary reparametrizations.
We apply this idea in the analysis of 1-loop corrections to the
AdS/CFT ansatz for the Wilson loop \cite{Maldacena:1998im,Rey:1998ik}. Our
main result is that for contours $C$ lying strictly on the boundary of
the AdS space, the 1-loop correction is finite, and the world-sheet
conformal invariance is preserved. On the contrary, for contours lying
inside AdS the reparametrization path-integral is logarithmically
divergent, and the corresponding amplitude is off shell and ill
defined.

Fixed contour amplitudes in flat space are off shell just as the
amplitudes for contours inside AdS. However, we decided to study the
corresponding reparametrization path-integral in great detail. One
reason for this study is to gain practical knowledge about
reparametrization path-integrals in general. The flat space background 
provides a good model example, because the action in this case is known
explicitly, and a rather precise analysis can be carried out.

A second reason comes from an observation that the $\sigma$-model
description of D0-branes gives rise to a formally equivalent
path-integral. This powerful analogy lead us to conjecture that the
corresponding non-local field theory of reparametrizations is
renormalizable, 
in the sense that all the divergences can be removed by adding counterterms
changing the shape of the contour. We checked this conjecture to the
2-loop order in the Feynman diagram expansion, and found in particular that
the divergences cancel provided the D0-brane equations of motion are satisfied.

There exists a natural generalization of the above result to the case of
higher dimensional branes, in which the role of reparametrizations is
played by maps from $S^1$ into the brane world-volume. 
We use this fact to find the first $\a'$ correction to the D$p$-brane
low energy effective action, and show that the result agrees with the
action computed from the $S$-matrix amplitudes.

The exact structure of the paper is as follows. In Section 2 we review
the general modern picture of gauge/string duality, discuss the string theory
ansatz for the Wilson loop, and explain the appearance of
reparametrization path-integrals. We show that the 1-loop correction
is divergent for contours in flat space, and finite for contours on
the boundary of AdS.
We also make a foray into the subject of loop equations,
reviewing old and recent work and discussing possibilities for future
research.

In Section 3 we study renormalization of the flat space
reparametrization path-integrals.
We calculate the 2-loop contour $\b$-function, and explain how it is
related to the D0-brane effective
action. Before treating the general case, we also
consider the circular contour example.

In Section 4 we deal with the D$p$-brane case. We compute a 2-loop
condition for cancellation of logarithmic divergences in a
corresponding path-integral, and interpret this condition as an
equation of motion following from a low energy effective action.

Section 5 is a short conclusion. Appendices A and B are
devoted to the details of Feynman diagram analysis. 

\newsection{Wilson loop in gauge/string duality}

\newsubsection{General picture}

The goal of the gauge/string duality program is to find a
string Lagrangian for 4d gauge theory color-electric flux lines. This Lagrangian must 
give permanent confinement at large distances, and at the same time
reproduce high-energy asymptotic freedom predictions.

It is well understood by now \cite{Gubser:1998bc,Polyakov:1998ju,Polyakov:2000fk} 
that for the pure Yang--Mills theory such ``confining strings'' must
propagate in a 5d gravitational background of the form
\be
ds^2=A(y)(dy^2+dx^2_{\!\mu}),\qquad A(y)\sim y^{-2}\qquad(y\to 0).
\label{*}
\ee
The gauge theory itself lives at the ``absolute'' $y=0$ of this space. 

The background \rf{*} will contain extra dimensions if the 4d gauge
theory in question has some extra matter fields, as it happens for the
much studied AdS/CFT example of the Yang--Mills theory with $\cN=4$
supersymmetry (see 
\cite{Aharony:1999ti} for a review). In this case
conformal symmetry fixes 
\be
A(y)=\frac{R^2}{y^2},
\label{conf}
\ee
so that the metric \rf{*} describes the $AdS_5$ space, while the full
background
is $AdS_5\times S^5$ due to the 6 scalars present in the 
supersymmetric Yang--Mills.

In physically interesting non-conformal cases the factor $A(y)$ will have logarithmic
corrections
\be
A(y)\sim \frac{R^2}{y^2}\Bigl[\log\Bigl(\frac{r_*}{y}\Bigr)\Bigr]^\a
\ee
near $y=0$, corresponding to the gauge theory logarithmic corrections to the
Coulomb law at short distances. Moreover, the space \rf{*} will
terminate at some finite $y$, leading to confinement. These features
have also been observed for gravity duals of strongly coupled gauge theories with
logarithmic RG flows \cite{Klebanov:2000hb}. 
 
Eventually, gauge/string duality must provide precise identification
of gauge invariant operators of the gauge theory with vertex operators of string theory
on the background \rf{*}, so that the corresponding field theory
correlators and string theory scattering amplitudes are equal. In the
standard AdS/CFT correspondence \cite{Aharony:1999ti} the gauge theory
coupling is strong, while the background \rf{*} is weakly curved, and
the supergravity approximation to the full string theory is
applicable.

For weakly coupled or asymptotically free gauge theories we cannot use
supergravity
and must solve the string $\s$--model with the world sheet action
\be
S=\frac 1{4\pi\a'}\int d^2\xi \sqrt{g}g^{ab}\d_aX^M\d_b X^N
G_{MN}(X)+\ldots
\ee
where $G_{MN}(X)$ is metric \rf{*} with $X=(y,x_\mu)$, and \ldots\
denotes
terms in the action corresponding to extra dimensions, world--sheet
fermions, and RR backgrounds needed to stabilize the space \rf{*}. In
conformal
cases there are explicit proposals for the full world--sheet action 
\cite{Polyakov:2000fk,Polyakov:2001af}. These $\s$--models have not
been solved so far.

\newsubsection{Wilson loop}

The gauge theory Wilson loop operators
\be
W[C]= \frac 1N\Bigl\langle\Tr{\rm Pexp}\oint_C A_\mu dx_\mu \Bigr\rangle_{\rm YM}
\ee
plays a special role in the gauge/string correspondence. In the dual
picture it should be described by an open string amplitude with
the string boundary tracing the contour $C$ \cite{Polyakov:1998ju}. In other words,
the Wilson loop must be given by a path-integral
\be
W[c_\mu(\s)]=\int [\cD X\cD g\ldots]\exp{-S[X,g\ldots]}
\label{W}
\ee
with the boundary conditions
\be
x_\mu(\s,0)=c_\mu(\s),\qquad y(\s,0)=0.
\ee

At present no general methods to work with \rf{W} are known. For
conformal theories \rf{conf} it is possible to consider 
the strong coupling limit
\be
\sqrt{\l}=\frac{R^2}{4\pi\a'}\gg 1\,.
\ee
In this case the integral \rf{W} can be computed in the classical
approximation as \cite{Maldacena:1998im,Rey:1998ik}
\be
W[c_\mu(\s)]\approx \exp -\sqrt{\l}\Am,
\label{class}
\ee 
where $\Am$ is the area of the minimal surface in $AdS_5$ bounded by
the contour $c_\mu(\s)$. World--sheet fermions and RR fields do not
contribute in this approximation.

As a first step towards a full theory, one can try to understand corrections
to \rf{class} perturbatively, as an expansion in $\l^{-1}$.
 
\newsubsection{Loop equation}

It has been known for a long time \cite{Makeenko:1979pb,Polyakov:1980ca,Polyakov:1987ez,Migdal:1983gj} that
gauge theory Schwinger--Dyson equations in the large $N$ limit give a
closed equation for the Wilson loop. This
``loop equation'' has the form
\be
\Lhat(\s)W[C]=W[C_1]*W[C_2],
\label{loop}
\ee
where the ``loop Laplacian'' 
\be
\Lhat(\s)=\lim_{\e\to 0}\int_{\s-\e}^{\s+\e}d\s'\,\frac{\gd^2}{\gd
c_\mu(\s)\,\gd c_\mu(\s')}
\ee
is the operator which picks up the $\gd$--function singular term in
the second variational derivative. The r.h.s. 
\be
W[C_1]*W[C_2]= -\l\oint_C\,\gd^{(4)}(y-c(\s))\,dy_\mu\,\dot{c}_\mu(\s)\, W[C_{1}]\,W[C_{2}]
\ee 
is nonzero only for $\s$ corresponding to a point of self-intersection, in which case it gets
contribution from the ``halves'' $C_1$ and $C_2$ obtained from $C$ by
splitting at that point. It is identically zero for
non-selfintersecting contours.

Strictly speaking, Eq.~\rf{loop} is derived in a regularized
theory. We would like to promote it to a well-defined equation for the
finite, renormalized Wilson loop. So far it has not been done, mainly
because it is unclear how to handle self-intersecting loops. If $C$ is
a smooth non-selfintersecting contour, then $W[C]$ has a finite
renormalized value \cite{Dotsenko:1980wb}. However, self-intersecting Wilson
loops
contain additional logarithmic divergences, and so do the loops with
corner points into which self-intersecting loops split \cite{Polyakov:1980ca,Brandt:1982gz}.

To avoid these difficulties, we may decide to
limit ourselves with non-selfintersecting loops. In this case it was
shown in \cite{Polyakov:2000ti} that the renormalized loop equation
\be
\Lhat(\s)W[C]=0
\label{loop1}
\ee
is well defined.

Eq.~\rf{loop1} by itself would not be sufficient to nonambiguosly
recover the Wilson loop functional. For example, it does not
distinguish between abelian and non-abelian gauge theories. Perhaps it
may be supplemented by some sort of ``boundary conditions'' for nearly
self-intersecting loops, which will make the identification unique. At
present this issue is not settled. However, even in the form \rf{loop1}
the loop equation may be used to test the gauge/string duality ansatz
\rf{W}.

For strongly coupled conformal gauge theory this test was performed in
\cite{Polyakov:2000ti,Polyakov:2000jg}, where we showed that the classical approximation \rf{class}
satisfies \rf{loop1} for any contour $C$ precisely in
the critical dimension $D=4$. This conclusion was reached by studying  
the second variational derivative of the nonlinear functional $\Am[C]$
in the short-distance limit $\s'\to\s$. We found the singular momentum
behavior of the form
\ba
\int\frac{\gd^2 \Am}{\gd c_\mu(\s-\frac h 2)\,\gd
c_\mu(\s+\frac h 2)}\;e^{iph}\,dh\hspace{6cm}\nonumber\\
=\ (1-D)|p|^3+(D-4)\,\ddot{c}^{\,2}(\s)\,|p|
+(4-D)\,\ddot{c}_\mu(\s)\,\frac{\gd\Am}{\gd c_\mu(\s)}\,p^0+\ldots\,,
\label{ptoinf}
\ea 
where \ldots\ denotes terms which are $O(p^{-1})$ for $p\to\infty$. 
The functional $\Am$ is invariant under reparametrizations
$c_\mu(\s)\to c_\mu(f(\s))$, and formula \rf{ptoinf}
is written in the natural gauge
$\dot{c}^{\,2}\equiv 1$.

In the coordinate space the first two terms in the r.h.s. of \rf{ptoinf} produce
non-local singularities proportional to $h^{-4}$ and $h^{-2}$, while
the third term corresponds to the local $\gd(h)$ singularity. The
loop Laplacian $\Lhat(s)W[C]$ is thus proportional to the coefficient
in front of the third term, and we see that indeed \rf{loop1} is satisfied provided that
$D=4$. 

\newsubsection{Quantum corrections\label{Quantum}}

As we move to smaller $\l$, quantum corrections to the classical
formula \rf{class} must become important\footnote{Although for the
Wilson loop in $\cN=4$ SYM formula \rf{class} is conjectured to hold 
to all orders in $\l$ if $C$ is a circle
\cite{Erickson:2000af,Drukker:2000rr}.}. At present the structure of
these corrections and their dependence on the contour are not
understood. In particular, I do not know if the loop equation
\rf{loop1} continues to hold when the corrections are taken into
account. If it turns out that it does not, this will perhaps 
have an interpretation based on the unavoidable presence of matter fields in conformal gauge theories
\cite{Polyakov:2000ti}. 

One--loop corrections to \rf{class} for type IIB string on
$AdS_5\times S^5$ described by the Green--Schwarz action were studied in
an important paper \cite{Drukker:2000ep} (see also
\cite{Forste:1999qn}). 
As it was shown in \cite{Drukker:2000ep}, 
logarithmic divergences in bosonic and
fermionic determinants coming from world--sheet oscillations
cancel. This important conclusion (related to the fact 
that $AdS_5\times S^5$ is an exact closed string
background) leaves however out the following puzzle. 

We do expect of course that the first quantum correction to \rf{class}
be finite, and conformal invariance in \rf{W} be preserved on the 1-loop level.
However, conformal invariance would be broken 
had we put contour $C$ at $y=\e>0$ rather than on the absolute. 

Consider for instance standard critical string theory in flat
space. It is well known that amplitudes with fixed boundary would be 
off shell, i.e. not conformally invariant, except for a very special
class of contours (straight lines, also known as D0-branes). 

Path--integral \rf{W} with a contour $C$ inside AdS must 
behave similarly to the flat space integral in this respect.
Only when the contour is pushed all the way to the absolute, where the
metric is singular, can we expect restoration of conformal invariance.
Apparently we are missing something, because the  
cancellation of bosonic and fermionic divergences observed in
\cite{Drukker:2000ep} does not seem to depend on
the precise position of the contour.

To resolve the puzzle, it is necessary to realize that besides the
world-sheet determinants, stringy Wilson loop \rf{W} receives a 1-loop
correction from oscillations of the classical action induced by
reparametrizations of the boundary contour. This is where
reparametrization path-integrals come into play. This correction must
be analyzed separately to decide if the conformal invariance is
preserved. In particular, it is the structure of this correction
that creates the difference between contours on the boundary of AdS on
the one hand, and contours in flat space and inside AdS on the other. 
Below we explain in detail how this happens, first in flat space, and then in AdS. 
 
\newsubsection{Flat space example}

Let us first recall what goes wrong with fixed
boundary amplitudes in flat space, considering bosonic string for
simplicity. We are to consider the path--integral
\be
Z=\int[\cD X^\mu \cD g]\,\exp(-{\textstyle\frac 1{4\pi\a'}}S_P),\qquad S_P=\int
d\s\,d\t\,\sqrt{g}\,g^{ab}\,\d_a X^\mu\, \d_bX^\mu,
\label{Z0}
\ee
with the boundary conditions
\ba
X^\mu(\s,0)=c^\mu(\s),
\label{bc}
\ea
where $c^\mu(\s)$ is a contour in flat $D$-dimensional space. 
The open string world-sheet with disk topology is parametrized 
by the upper half-plane.

In conformal gauge the action becomes
\be
S=\int d\s\, d\t\, (\d X)^2\,.
\ee
World-sheet oscillations and ghosts give rise to determinants that
cancel each other provided that $D=26$. The answer we seem to be getting is
\be
Z\stackrel{?}{=} \exp(-{\textstyle\frac 1{4\pi\a'}}\Scl[c_\mu(\s)])\,,
\label{Z1}
\ee
where $S_{\rm cl}$ is the classical action for
the solution of the Dirichlet problem on the upper half-plane with the
given boundary conditions
\be
S_{\rm cl}=\int \frac{dp}{2\pi}\,|p|\,c^\mu(p)\,c^\mu(-p)\,.
\label{scl}
\ee
However, Eq.~\rf{Z1} is not, cannot be correct. This can already be
seen from the fact that in the classical  $\a'\to0$ limit \rf{Z0} must be
dominated by the surface with minimal area rather than minimal energy.
Even more, \rf{Z0} is formally invariant under reparametrizations
$\s\to\a(\s)$, while the classical Dirichlet action $\Scl$ is not.

The correct answer is  
\be
Z=\int[\cD \a(\s)]\exp(-{\textstyle\frac
1{4\pi\a'}}\Scl[c_\mu(\a(\s))])\,,
\label{Z2}
\ee
where the path-integral is over the group of boundary reparametrization.
This expression is manifestly reparametrization invariant (modulo
possible quantum anomalies). It also possesses the right 
classical limit, because 
\be
\min \limits_{\{\a(\s)\}} \Scl[c_\mu(\a(\s))]=minimal\ area\,.
\label{min}
\ee
The reason why \rf{Z2} rather than \rf{Z1} follows from \rf{Z0} is
that fixing conformal gauge on the disc is in general
impossible unless we allow diffeomorphisms 
reparametrizing the boundary. After we trade the integral over metrics
for an integral over $\Diff\times Weyl$, the part $\Diff_0\times Weyl$
produces but an infinite volume factor, while the $\Diff(S^1)$ component gives
rise to \rf{Z2}. More details can be found in Appendix A.

Eq.~\rf{Z2} has appeared in the literature before, see
e.g. \cite{Polyakov:1987ez,Cohen:1986sm},
but the dynamical consequences of this representation to the best of
my knowledge have not been explored. 

Since we expect that for a general contour 
conformal invariance in \rf{Z0} is broken, the integral \rf{Z2}
must contain logarithmic divergences, and it indeed does.
Let us see how this happens in the 1-loop approximation. Without loss of generality, we may
assume that $\a_*(\s)\equiv\s$ provides minimum in
\rf{min}\footnote{Otherwise we have to reparametrize $c_\mu(\s)\to
c_\mu(\a_*(\s))$.}. The action in coordinate space becomes
\be
\Scl[c_\m(\s+\b(\s))]=-\,\frac 1\pi \int
d\s\,d\s'\,\frac{c_\m(\s+\b(\s))\, c_\m(\s'+\b(\s'))}{(\s-\s')^2}.
\label{becomes}
\ee 
The part of the action quadratic in $\b(\s)$ is given by
\be
\St=-\,\frac 1\pi \int
d\s\,d\s'\,\frac{\dot{c}_\m(\s)\,\dot{c}_\m(\s')}{(\s-\s')^2}\,\b(\s)\,\b(\s')+
\frac 12\int d\s\,\frac{\gd\Scl}{\gd c_\m(\s)}\,\ddot{c}_\mu(\s)\,\b^{\,2}(\s),
\label{s2}
\ee
where
\be
\frac{\gd\Scl}{\gd c_\m(\s)}=
-\,\frac 2\pi\int
d\s'\,\frac{c_\m(\s')}{(\s-\s')^2}\,.
\ee 
For later reference notice that we have
\be
\frac{\gd \Scl}{\gd\b(\s)}=\frac{\gd\Scl}{\gd
c_\m(\s)}\,\dot{c}_\m(\s)\equiv 0,
\label{db}
\ee
since the linear in $\b(\s)$ term in the action has to vanish.

The first quantum correction in \rf{Z2} corresponds to the
path-integral
\be
\int [\cD \b]\,\exp(-\St).
\ee
This integral contains a logarithmic divergence. Indeed, from the
first term in $\St$ we see that the propagator of the field $\b(\s)$
in the mixed representation behaves like
\be
\langle \b(p) \b(-p) \rangle = \frac{1}{2|p|\dot{c}^2(\s)}+O(p^{-2})\qquad(p\to\infty).
\ee
It follows that 
\be
\langle \b^{\,2}(\s) \rangle = \frac{\ll}{2\pi\dot{c}^2(\s)}+finite,
\ee
where $\L$ is the momentum cutoff.

Since the second term in $\St$ involves $ \b^{\,2}(\s)$, we get a
logarithmically divergent contribution to the effective action
$F=-\log Z$
\be
F_{\rm div}=\frac{\ll}{4\pi}\int d\s\, 
\frac{\gd\Scl}{\gd c_\m}\,\frac{\ddot{c}_\m}{\dot{c}^2}\,.
\label{logdiv}
\ee
The same divergence can be derived in a different way, by using 
heat-kernel methods to analyze the determinant of the integral operator
corresponding to $\St$. 

It is worth noting that the divergence \rf{logdiv} is non-local and cannot be removed by adding
local counterterms to the contour action. Conformal invariance would be indeed
broken unless it canceled. The most general local condition for this
to happen is 
\be
\ddot{c}(\s)\equiv 0\,.
\label{straight}
\ee
This means that the contour has to be a straight line for \rf{Z2} to
be well defined. The straight line boundary condition would be
identical to having a D0-brane, if it were not for a subtle
difference \cite{Polyakov:1998tj}. Namely, in \rf{Z2} the surface is
attached to the contour without folds, while for the D0-branes the folds
are allowed. 

I will have a lot more to say about the path-integral \rf{Z2} and its
relation to the bosonic D-branes in Sections 3 and 4 of this paper. Now let us return
to the Wilson loop in AdS.

\newsubsection{Wilson loop in AdS is 1-loop finite}

In the AdS case, a 1-loop analysis very similar to the one given
in the previous section can be carried out. As I mentioned in Section 2.4, 
world-sheet oscillations do not lead to logarithmic
divergences in this order. As a result, all possible troubles are
connected with the AdS analogue of \rf{Z2}\footnote{This integral over
reparametrizations was disregarded in \cite{Drukker:2000ep}.}
\be
\int[\cD \a(\s)]\exp(-{\textstyle\frac
1{4\pi\a'}}\Sa[c_\mu(\a(\s))]).
\label{ads}
\ee
Here $\Sa[c_\m(\s)]$ is the classical Dirichlet action in the AdS space
\be
\Sa=\min \int d^2\xi\,\frac{(\d x_\mu)^2+(\d y)^2}{y^2}
\ee
with boundary conditions \rf{bc}.

Unlike in flat space, no explicit formula similar to \rf{scl} exists
for $\Sa$. However, to find the 1-loop divergences, it is sufficient to
understand short-distance singularities of the quadratic action. In
other words, we have to study
\be
\frac{\gd^2\Sa}{\gd\b(\s)\,\gd\b(\s')}= 
\frac{\gd^2\Sa}{\gd c_\mu(\s)\,\gd c_\nu(\s')}\,\dot{c}_\mu(\s)\,
\dot{c}_\nu(\s')+ \frac{\gd\Sa}{\gd c_\mu(\s)}\,\ddot{c}_\mu(\s)\,\gd(\s-\s').
\label{bb}
\ee  
The latter formula folows by Taylor expanding $c_\mu(\s+\b(\s))$.

Short-distance limit of the second derivative of $\Sa$ was analyzed in
\cite{Polyakov:2000jg}, where we found that
\be
\int\frac{\gd^2 \Sa}{\gd c_\mu(\s-\frac h 2)\,\gd
c_\mu(\s+\frac h 2)}\;e^{iph}\,dh =
\Bigl(3\,\frac{\dot{c}_\mu\,\dot{c}_\nu}{\dot{c}^4}-\frac{\gd_{\mu\nu}}{\dot{c}^2}\Bigr)|p|^3+const\,\frac{\dot{c}_{[\mu}\,\ddot{c}_{\nu]}}{\dot{c}^4}\,p^2+O(p).
\label{cc}
\ee 
All coefficients in the r.h.s. are evaluated at $\s$. 
It is crucial for what I am going to say 
that the $p^2$ term is multipled by an antisymmetric tensor (in fact,
this follows from $\s\to-\s$ symmetry of the problem). 
After substituting into \rf{bb}, the $p^2$ term vanishes, and we obtain
\be
\int\frac{\gd^2 \Sa}{\gd \b(\s-\frac h 2)\,\gd
\b(\s+\frac h 2)}\;e^{iph}\,dh = 2|p|^3+O(p).
\label{bb1}
\ee
This relation is sufficient to show that the path-integral over
$\b(\s)$ will be free from 1-loop logarithmic divergences. From the
most singilar term we read off the propagator
\be
\langle \b(p) \b(-p) \rangle = \frac{1}{2|p|^3}\,.
\ee
Possible vertices coming from lower order terms in \rf{bb1}
will produce integrals with a UV asymptotic behavior of
\be
\int^\infty dp\,\frac{O(p)}{|p|^3}\,.
\ee
Since this is absolutely convergent, logarithmic divergences will be absent. 
Again, the same conclusion can be reached by heat-kernel analysis
departing from \rf{bb1}.

The above derivation resolves the puzzle mentioned in Section 2.4 and 
explains quantitatively the difference between
having the contour on the absolute and inside the AdS. For a contour
strictly inside, the dependence of the action
on reparametrizations would be of the form $|p|+O(p^0)$ similar to
\rf{s2}, and conformal invariance would be broken by logarithmic
divergences.  

\newsection{Fixed contours and D0-branes}

As we saw in the previous section, fixed boundary amplitudes in string
theory lead naturally to path-integrals over reparametrizations after
conformal gauge fixing. These integrals are sure to contain important
dynamical information relevant to gauge/string duality, and it is very
important to learn how to work with them.

As a first step in this direction, we are going to perform a detailed
analysis of the flat-space integral \rf{Z2}. One should not forget of course that our
main interest is with the AdS integral \rf{ads}. However, its treatment
is greatly complicated by not knowing the action explicitly, and I
postpone it to a later occasion. The integral \rf{Z2} has a definite
advantage in this respect. We will be able to obtain rather precise
statements about it, and along the way learn important general
lessons about the nature of reparametrization path-integrals.

\newsubsection{D0-brane analogy}

The flat-space ``Wilson loop'' integral \rf{Z2} is formally identical
to a path-integral arizing in the $\s$-model description of a bosonic
D0-brane, when only the fields describing the shape of the brane are
turned on. To see this, it is sufficient to rewrite the usual D0-brane
boundary conditions
\be
X^0|_{\t=0}=free,\qquad X^i|_{\t=0}=\f^i(X^0)\qquad(i=1\ldots25)
\ee
in the covariant form
\be
X^\m(\s,0)=c^\m(\a(\s)).
\label{single}
\ee
As a result, we get an integral of exactly the same form as \rf{Z2}.

In this approach the only difference between Wilson loops and
D0-branes is in the nature of the field $\a(\s)$. In the Wilson loop    
path-integral, $\a(\s)$ is a diffeomorphism so that
\be
\dot{\a}(\s)>0\,.
\ee
However, in the D0-brane case $\a(\s)$ does not have to satisfy this
constraint, because the field $X^0$ was free to backtrack. In other
words, the worldsheet attached to the D0-brane can have folds. The measure
of integration is also different
\ba
\|\gd\a\|^2\is\int d\s\,[\gd\a(\s)]^2\hspace{1.8cm}(\mbox{D0-brane}),\\
\|\gd\a\|^2\is\int d\s\,\dot{\a}(\s)\,[\gd\a(\s)]^2 \qquad(\rm{Wilson\  loop}).
\ea
The former is the usual Gaussian measure, while the latter is the
measure on $\Diff(S^1)$ invariant under right shifts
$\a(\s)\to\a(f(\s))$.

The group $\Diff(S^1)$ has a boundary consisting of $\a(\s)$ for which
\be
\dot{\a}(\s)=0
\label{diffbound}
\ee
on an interval. 
I am quite sure that this boundary affects nonperturbative dynamics,
and in particular gives rise to an anomaly in reparametrization Ward
identities. However, at present I am unable 
to demonstrate this quantitatively. On the other hand, I expect and
assume for the purposes of this paper that the
perturbation theory of small oscillations around a particular
$\a_*(\s)$ should not feel the presence of the boundary \rf{diffbound}. 

Admittedly, this assumption has to be looked into more
carefully. However, even if it will later be found to be incorrect,
this will not invalidate what I have to say below, but rather just
restrict it to the D-brane case. Such a discovery of a different
perturbative sector in critical string theory 
seems an exciting however unlikely possibility.
  

\newsubsection{Renormalization and D0-brane effective action}

From a practical point of view, the path-integral \rf{Z2} describes a
nonlocal quantum field theory of the field $\a(\s)$. As we saw in
Section 2.5, in spite of being 1-dimensional, this field theory is
UV divergent (unlike say the usual quantum mechanics).

In Section 2.5 we concluded that if the contour is a straight line, 
the 1-loop divergence \rf{logdiv} is absent. For a general
contour this divergence is nonlocal and cannot be removed by local
counterterms. However, there is another natural way to deal with it. 
Namely, since the divergence is proportional to $\gd\Scl/\gd c$, it
 can be removed by renormalizing the contour
\be
c\to c+\gd c,\qquad \gd
 c^\m=-\a'\ll\,\frac{\ddot{c}^\m}{\dot{c}^2}\,.
\ee
In fact because of \rf{db} I can add any multiple of $\dot{c}^\m$ to
 this formula. It is natural to write this 1-loop RG in the
 reparametrization-invariant form
\be
\b^\m(c)=\frac{dc^\m}{d\ll}=-\a'\,\frac{\ddot{c}^\m_\perp}{\dot{c}^2}\,,\qquad
\ddot{c}^\m_\perp=\ddot{c}^\m-\dot{c}^\m\frac{(\dot{c}\,\ddot{c})}{\dot{c}^2}
\,.
\label{RG}
\ee
The ``$\b$-function equation''
\be
\b^\m=0
\ee
is equivalent to \rf{straight} and has straight lines as solutions.

It is well known \cite{Callan:1985ia,Polyakov:1987ez} that string
theory world-sheet $\b$-functions are related to the low energy
effective action by a generic formula
\be
\b^m(\f)=g^{ml}(\f)\,\frac{\gd\G(\f)}{\gd\f^l}\,,
\label{Gamma}
\ee
where $g^{ml}(\f)$ is a nondegenerate metric in the space of massless
space-time fields $\f$..

A direct way to compute $\G$ would be to calculate $S$-matrix amplitudes and construct a
space-time action which reproduces them. Formula \rf{Gamma} says that
instead we may consider the nonlinear $\s$-model describing a string propagating in background
fields $\f$. Conditions of conformal invariance of this $\s$-model
will coincide with equations of motion for $\G$. 
This truly remarkable equivalence holds in all known
cases, although a general proof to the best of my knowledge has not
been given.

The advantage of the second method is that it produces a manifestly
covariant derivative expansion of $\G$. To obtain higher and higher
order in $\a'$ terms in this expansion, one just has to compute the
$\s$-model $\b$-function to more and more loops.

It is reassuring to see that our RG \rf{RG} can be rewritten in the
form of \rf{Gamma} as
\be
\b^\m(c)=-\,\frac{\a'}{\sqrt{\dot{c}^2}}\,\frac{\gd\Seff}{\gd
c^\mu(\s)}\,,
\label{relation}
\ee
where
\be
\Seff=\int d\s\,\sqrt{\dot{c}^2}\,.
\label{D0}
\ee
Thus we recover the usual bosonic D0-brane effective action (in the
particular background $G_{\m\n}=\eta_{\m\n}$, $B_{\m\n}=0$, $\F=const$).
We see that the reparametrization path-integral provides a convenient
framework for obtaining this classical result. 

While the 1-loop calculation can be actually done directly on the
world-sheet without invoking the non-local action \rf{scl} explicitly
(see \cite{Leigh:1989jq}), the representation \rf{Z2} becomes essential when we move
beyond one loop, as we do below.

\newsubsection{Renormalizability beyond one loop}

Let me repeat the logic of the above discussion. We considered a
1-dimensional QFT with partition function \rf{Z2}, and found that it
is 1-loop renormalizable. At first this may seem like a pure
coincidence, since the theory is non-local, and no general
renormalizability arguments apply to it. However, the relation with
D0-brane dynamics expressed by \rf{relation} suggests that this is not
so. In fact, I conjecture that renormalizability must hold to any loop
order. Moreover, the $\a'$ expansion of the $\b$-function must have
the form following from \rf{Gamma}
\be
\b^\mu(c)=g^{\mu\nu}(c)\,\frac{\gd\Seff}{\gd
c^\n(\s)}\,.
\ee
The metric $g^{\m\n}(c)$ may and will in general get $\a'$ corrections compared
to \rf{relation}
\be
g^{\m\n}(c)=-\,\frac{\a'}{\sqrt{\dot{c}^2}}\,\gd^{\mu\nu}+\ldots
\ee
As for the $\Seff$, the only possible covariant corrections must be
constructed from higher order derivatives
with respect to the natural parameter $s$, such as the curvature
$c''_{ss}$, torsion $c'''_{sss}$, etc. However, all these terms would vanish on shell, that is for
straight lines. Thus we arrive at the conclusion that the effective
action \rf{D0} should not get any perturbative corrections.

Two-loop computations that I do below confirm these predictions about
renormalizability and the form of the $\b$-function.

\newsubsection{Circular contour}

Before we tackle 2-loop corrections for a general contour, it is
instructive to consider the example when $C$ is a circle
\ba
c^0(\s)\is R\cos\s\,,\nonumber\\  
c^1(\s)\is R\sin\s\,,\nonumber\\
c^i(\s)\is 0\qquad(i=2\ldots 25)\,.
\ea
In this case the only renormalizable parameter is the radius $R$. It
is convenient to include it into the definition of the coupling
constant $g$
\be
\frac{1}{g}=\frac{R^2}{4\pi\a'}\,. 
\ee
Because we use $0<\s<2\pi$ to parametrize the circle, the classical
action \rf{scl} changes to the discrete sum  
\be
\Scl=\frac1{2\pi}\sum_{p\in\Z}|p|\,c^\m_{p}\,c^\m_{-p}\,.
\ee
The action $\Scl[c^\m(\a(\s))]$ is easily found by using the
representation
\be
c^\m(\a(\s))\,c^\m(\a(\s'))=\mbox{Re}\, \exp i(\a(\s)-\a(\s'))\,.
\ee
It is given by
\be
\Scl[c^\m(s+\b(\s))]=\frac
1{4\pi}\sum_{p\in\Z}(|p+1|+|p-1|)\sum_{k,l\ge0}
\frac{i^{k-l}}{k!\,l!}\,(\b^k)_p\,(\b^l)_{-p}\,.
\label{sbeta}
\ee
It is easy to see that all odd order terms vanish in this action. In particular, the linear
term in $\b$ is absent. The quadratic action is given by
\be
S_2=\frac1{4\pi}\sum
E(p)\,\b_p\,\b_{-p}\,,\qquad E(p)=|p+1|+|p-1|-2\,.
\ee
It should come as no surprise that $E(p)=0$ for $p=0,\pm1$. These
three zero modes reflect the $SL(2,\RR)$ invariance of $\Scl$.

We view \rf{Z2} as a non-local quantum theory of the field
$\beta(\s)$ with the classical action \rf{sbeta}. We expect that all
correlators of this theory can be made finite by adding
counterterms to the action renormalizing the coupling constant $g$. 
Below I will show that this is indeed true for all 4-point functions in the
1-loop order, and for all 2-point functions in the 2-loop order. 
We will also find the 2-loop $\b$-function.

To perform this calculation, we will need the first two higher order
terms in \rf{sbeta}
\ba
S_4\is\frac1{2\pi}\sum
E(p)\,\Bigl(\frac 18\,(\b^2)_p\,(\b^2)_{-p}-\frac16\,\b_p\,(\b^3)_{-p}\Bigr)\,,\nonu
S_6\is\frac1{2\pi}\sum
E(p)\,\Bigl(\frac1{120}\,\b_p\,(\b^5)_{-p}-\frac
1{48}\,(\b^2)_p\,(\b^4)_{-p}+\frac
1{72}\,(\b^3)_p\,(\b^3)_{-p}\Bigr)\,.
\ea
Denoting the nonlocal $E(p)$ vertex by a wavy line, we have the
following diagrammatic representation
\ba
S\is S_2+S_4+S_6+\ldots\nonu
\is\frac 12\;\fig{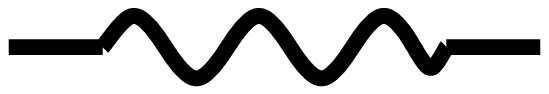}{0.17}{1}+
\Bigl(\frac18\;\fig{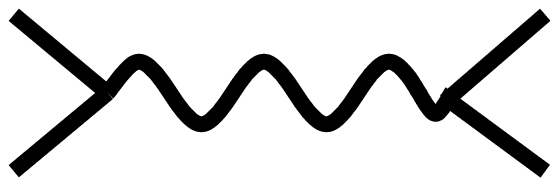}{0.17}{-2}
-\frac16\;\fig{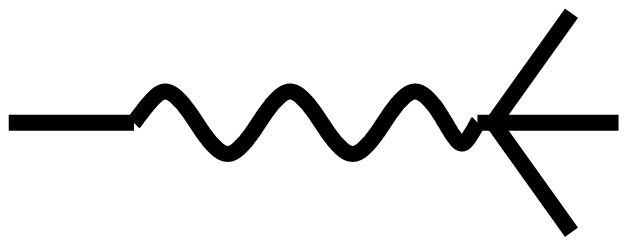}{0.17}{-3}\Bigr)\nonu
&\!\!+\!\!&\Bigl(\frac1{120}\;\fig{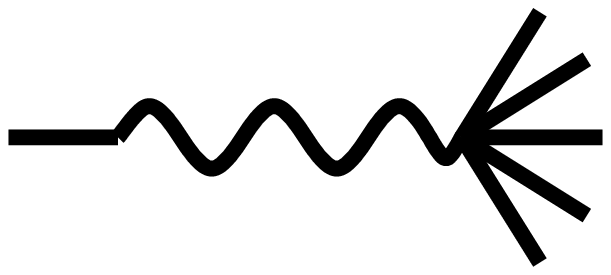}{0.17}{-4}
-\frac1{48}\;\fig{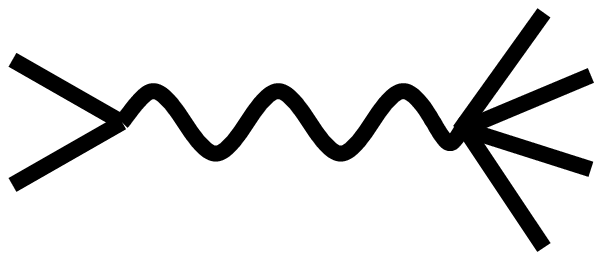}{0.17}{-4}+
\frac1{72}\;\fig{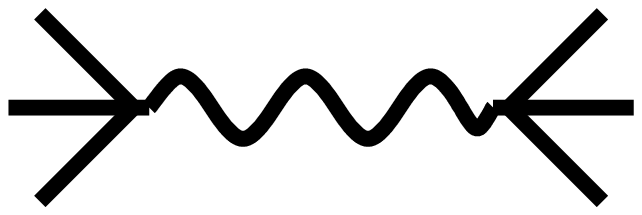}{0.17}{-2}\Bigr)+\ldots\,.
\label{orig}
\ea
Contracting these diagrams with the propagator
$E(p)^{-1}$, we obtain the 1-loop correction to the quantum effective
action through $\b^4$ terms 
\ba
S_{1\mbox{-}loop}\is g\Bigl(\frac 12\;\fig{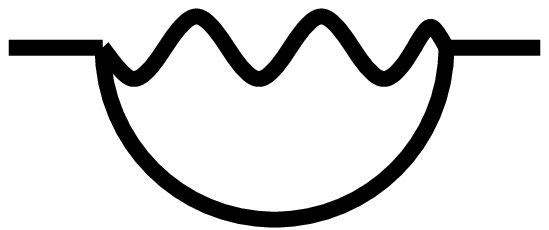}{0.17}{-6}
-\frac 12\;\fig{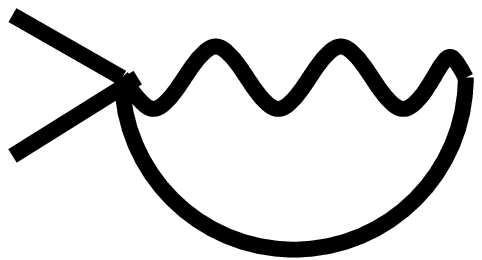}{0.17}{-6}
-\frac 12\;\fig{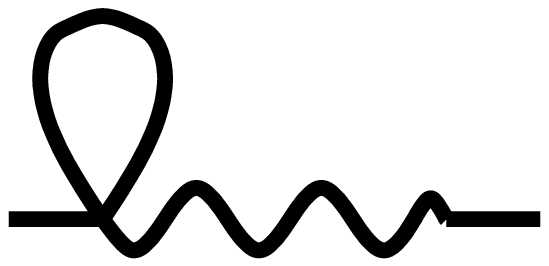}{0.17}{0}\nonu
&&+\,\frac1{12}\;\fig{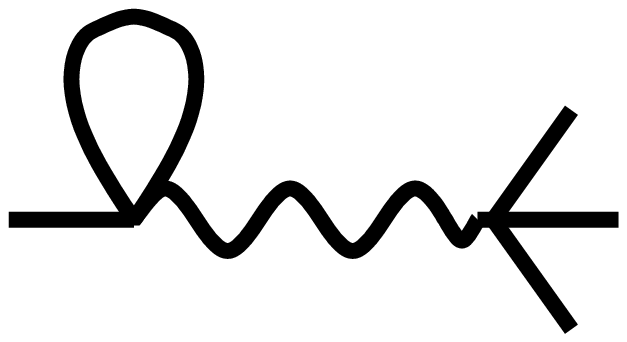}{0.17}{-2}
-\frac18\;\fig{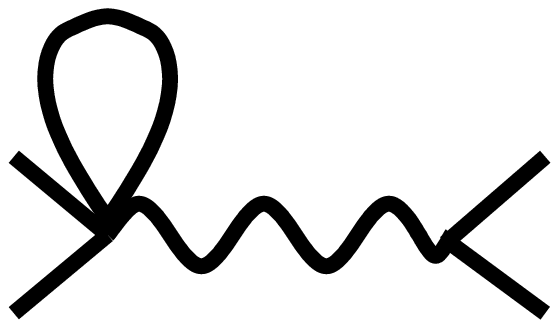}{0.17}{-2}
+\frac1{12}\;\fig{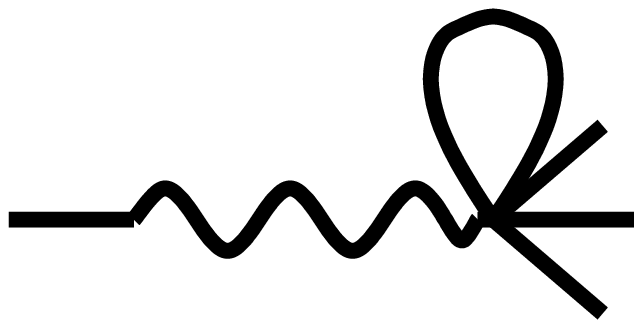}{0.17}{-2}\nonu
&&+\,\frac1{24}\;\fig{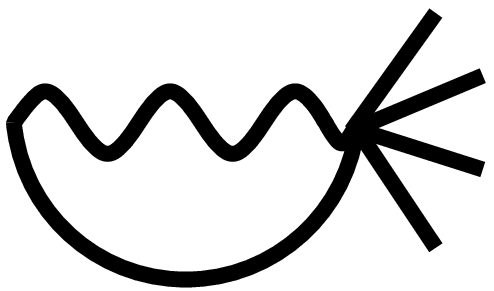}{0.17}{-6}
-\frac16\;\fig{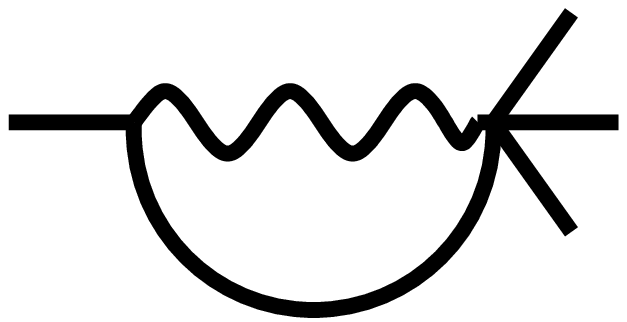}{0.17}{-6}
+\frac18\;\fig{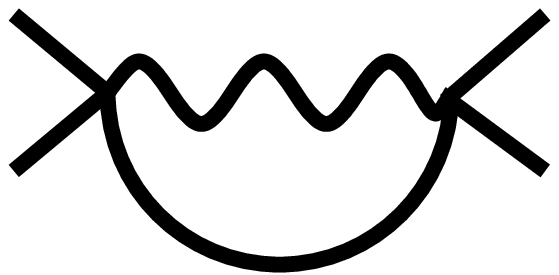}{0.17}{-6}\Bigr)+\ldots\,.
\label{as}
\ea
Notice that the diagrams like \fig{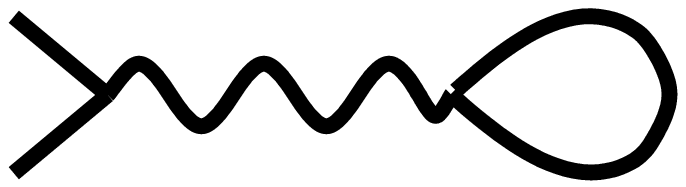}{0.17}{-3} vanish, because $E(0)=0$.
The change in the numerical factors reflects the numbers of equivalent contractions.

The r.h.s.~of \rf{as} is divergent. First of all, it contains logarithmic divergences
proportional to the ``leaf'' diagram
\be
  \fig{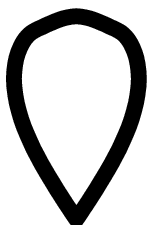}{0.17}{-2}\,=\frac1{2\pi}\sump_{|p|<\L}\frac1{E(p)}=\frac1{2\pi}\ll+finite,
\ee
where $\sump$ is being taken over $p\ne0,\pm1$.
Superficially, it also contains linear divergences proportional to the ``oyster''
\fig{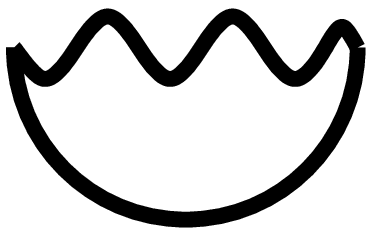}{0.17}{-2}. 

However, as the reader may easily check, the ``oyster'' divergences of
individual diagrams are local and cancel
each other when added (for instance, in the last line of \rf{as} this
happens because $1/24-1/6+1/8=0$).
What remains is the overall logarithmic divergence, and it is easy to
see that it is exactly proportional to the $S_2+S_4$ part 
of the classical action. It can be removed by adding the counterterm
\be
S_{ct}=\Bigl(\frac {g}{2\pi}\ll\Bigr) S\,
\ee
which translates into the RG law for the bare coupling constant
\ba
\frac1{g(\L)
}\is\frac 1 g +\frac 1{2\pi}\ll\,,\label{glam}\\[4pt]
\b(g)\is\frac {dg}{d\ll}=-\frac {g^2}{2\pi}\,.\label{betag}
\ea
A simple check shows that this 1-loop $\b$-function agrees with the
general contour result \rf{RG}.

It is interesting to note that the partition function renormalization
of Section 3.2 and the above correlator renormalization follow in fact
from two quite different and formally inequivalent computations. Their
agreement should not be taken as a pure coincidence. Rather, it is a sign of
a renormalizable field theory structure hidden behind.

Let us now discuss order $g^2$ corrections to the quantum effective
action. In this order 2-loop terms coming from the original expansion
\rf{orig} mix with contributions of 1-loop counterterms. The quadratic
part of the effective action consists of the following diagrams
\be
\begin{array}{ll}
\mbox{(a)}\ \makebox[0pt][l]{\fig{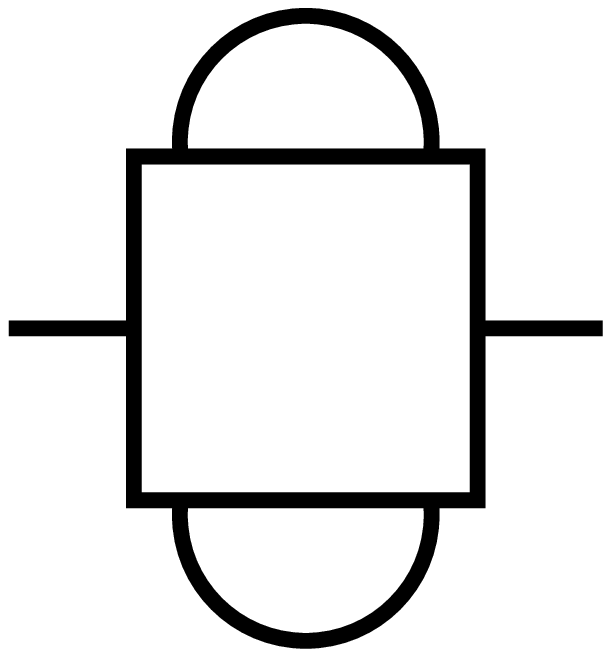}{0.17}{-13}}\hspace{10pt}S_6\ \ \ -
\frac 12\; \makebox[0pt][l]{\fig{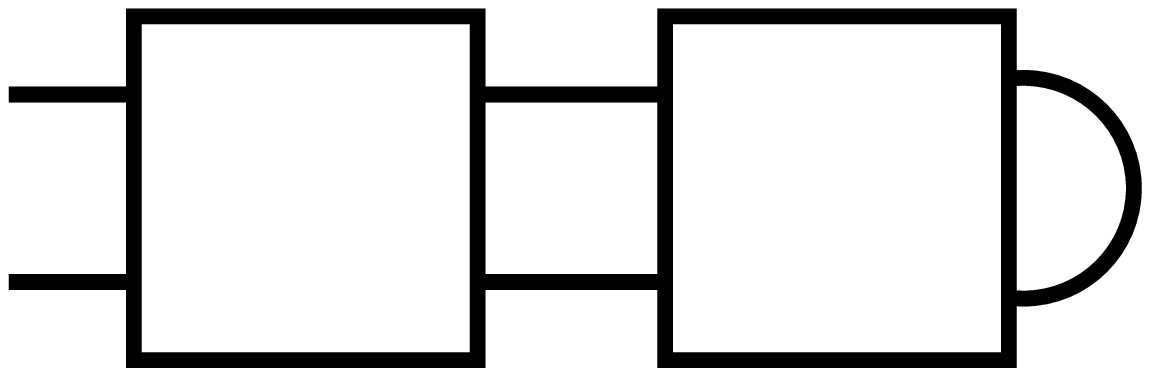}{0.17}{-6}}\hspace{9pt}S_4\hspace{14pt}S_4\qquad\qquad
&
\mbox{(b)}\ -\, \frac 12\; \makebox[0pt][l]{\fig{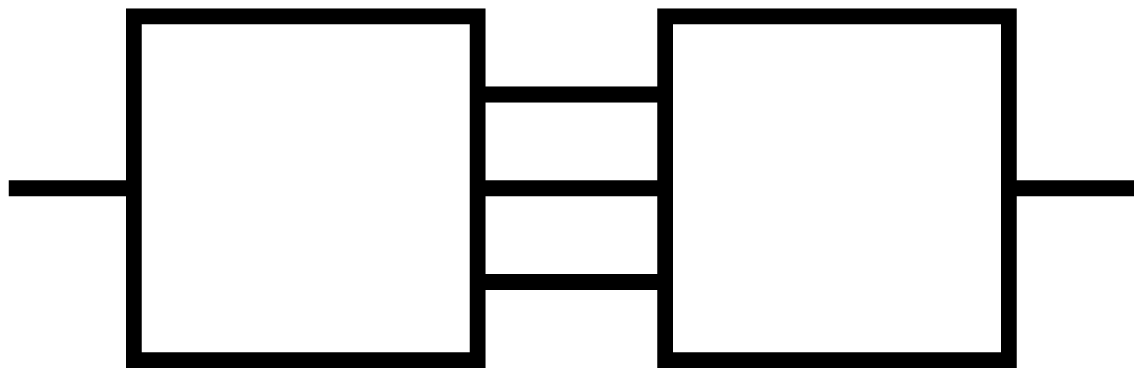}{0.17}{-6}}\hspace{9pt}S_4\hspace{14pt}S_4\\[20pt]
\mbox{(c)}\
\makebox[0pt][l]{\fig{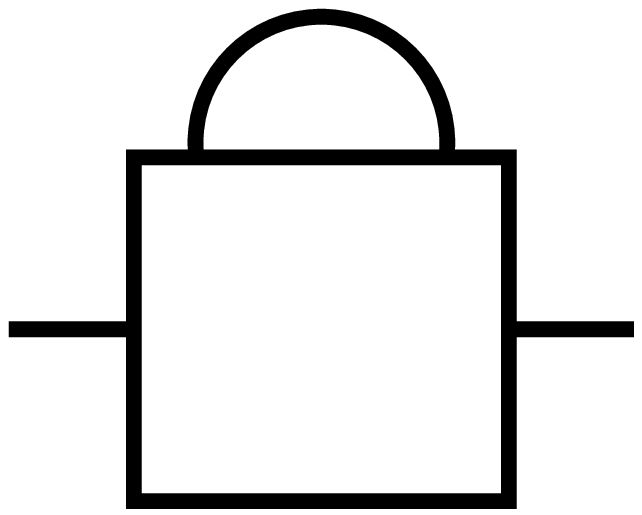}{0.17}{-6}}\hspace{9pt}S_4^{ct}\ \ \; 
-\;\makebox[0pt][l]{\fig{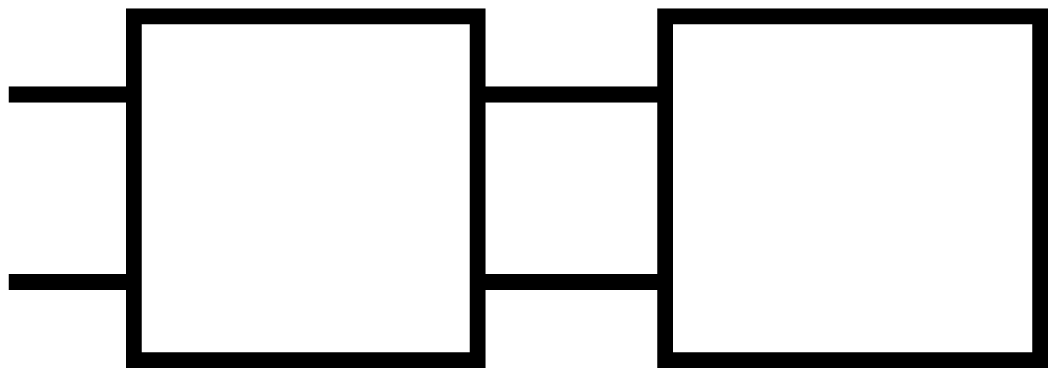}{0.17}{-6}}\hspace{9pt}S_4\hspace{14pt}S_2^{ct}
&\mbox{(d)}\ -\, \frac 12\;  \makebox[0pt][l]{\fig{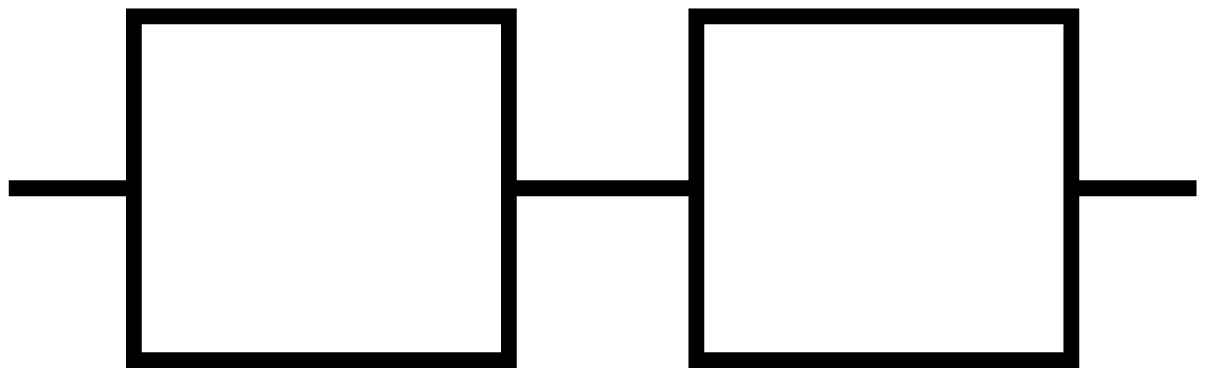}{0.17}{-6}}\hspace{9pt}S_2^{ct}\hspace{13pt}S_2^{ct}
\end{array}
\label{diver1}
\ee
It is quite easy to see directly that the divergent parts of the diagrams in (c)
 cancel identically. The same happens for (b), although in a subtler way: 
the diagrams fall into groups, and within each
group divergences cancel. For instance, one of the groups is
\be 
\frac12\fig{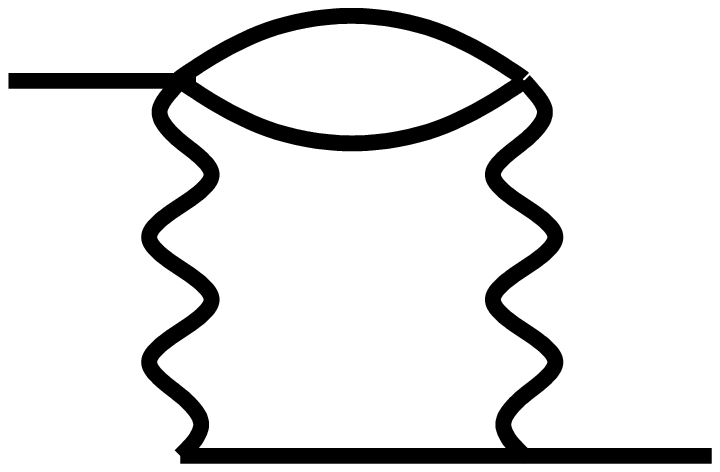}{0.17}{-15}-\frac14\fig{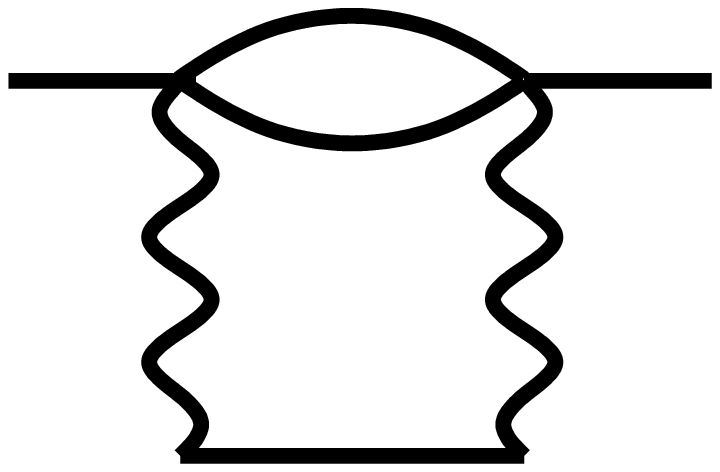}{0.17}{-15}-\frac14\fig{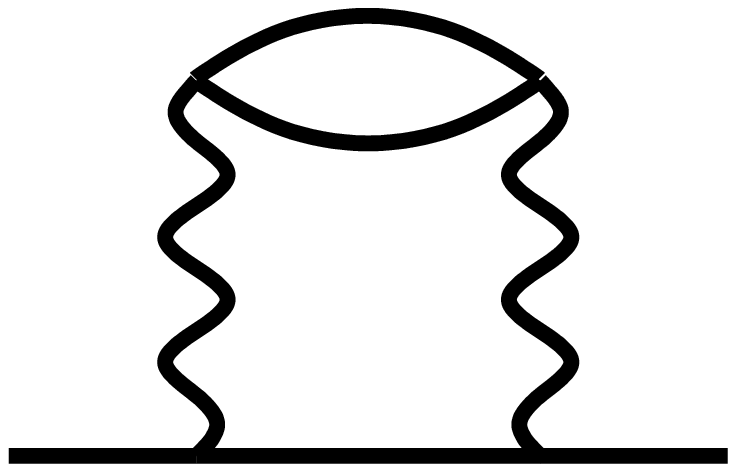}{0.17}{-15}<\infty\,.
\label{group}
\ee
Class (a) is the most numerous one, it consists of about 20
diagrams. Again many cancellations similar to \rf{group} occur. In the
end all that remains are two diagrams from the first term
\be
\frac 18\fig{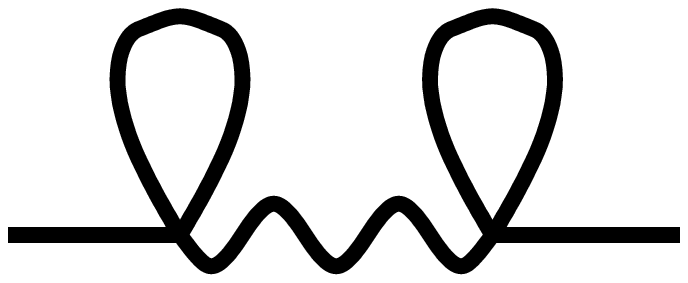}{0.17}{-15} +\frac 18\fig{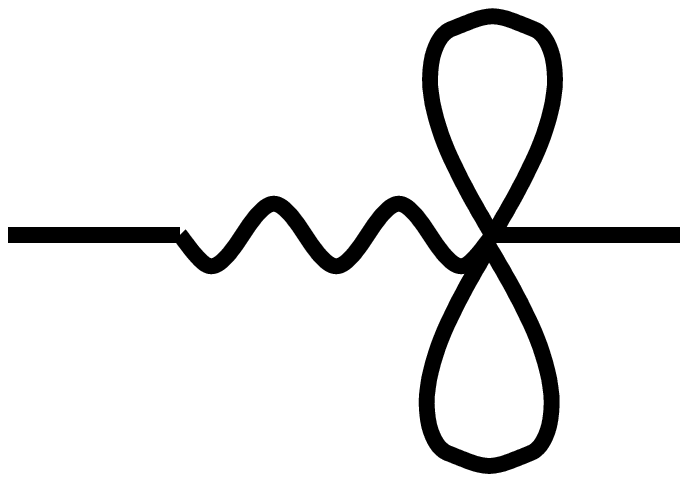}{0.17}{-15}
\label{diver2}
\ee
and three diagrams from the second one
\be
\frac 12\fig{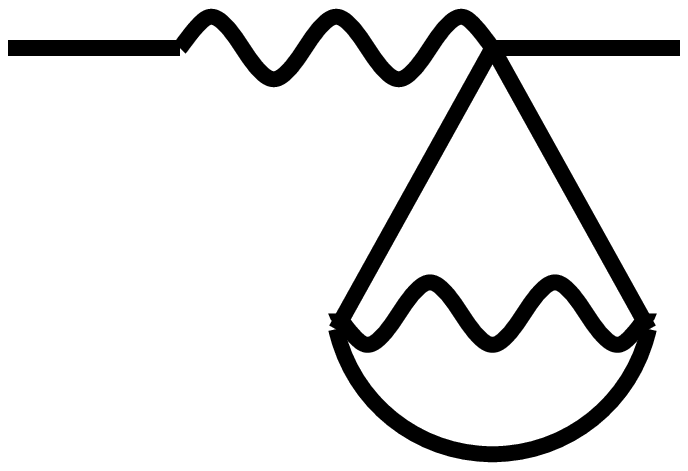}{0.17}{-25}-\frac 12\fig{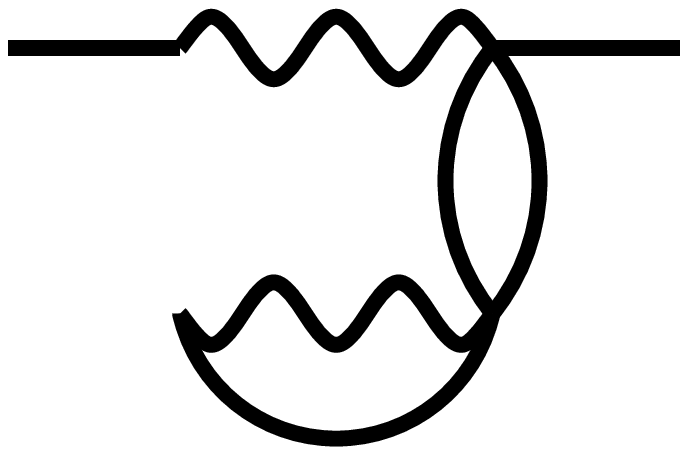}{0.17}{-25} -\frac12\fig{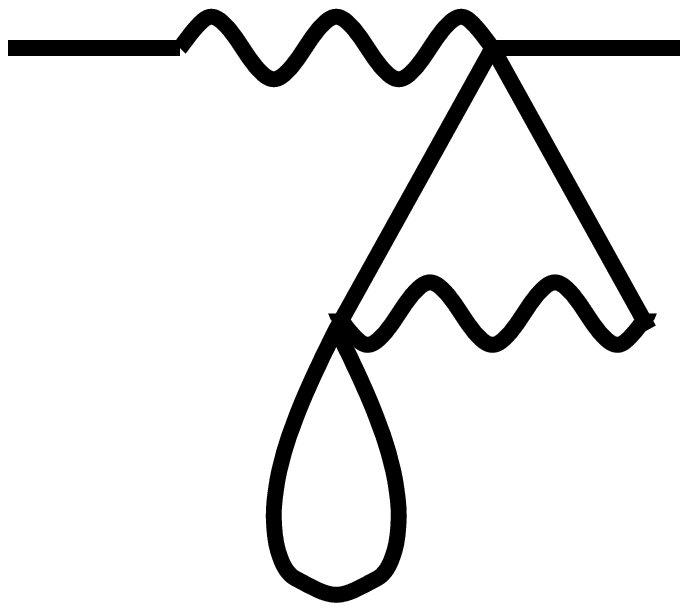}{0.17}{-25}\,.
\label{diver3}
\ee
The only nontrivial calculation is required for the divergence
\ba
\fig{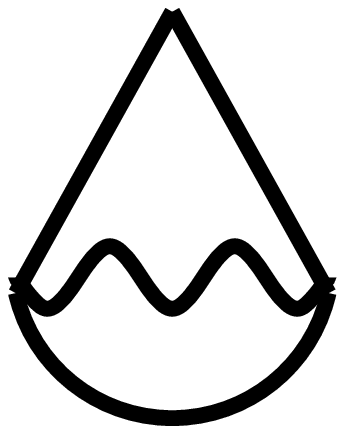}{0.17}{-15}-
\fig{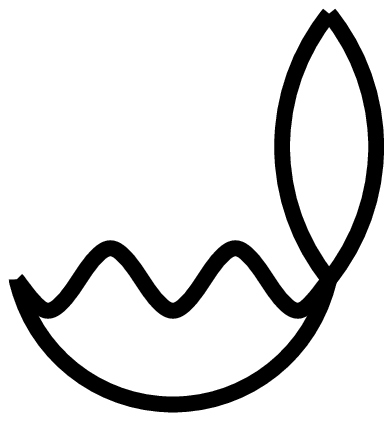}{0.17}{-15}\is\frac{1}{(2\pi)^2}\sump_{|q|,|p|<\L}\frac{E(q-p)-E(q)}{E(p)^2
E(q)}\nonu
\is\,\frac1{8\pi^2}(\ll)^2+\frac1{4\pi^2}(\g-1)\ll + finite\,,
\ea
where the Euler constant $\g$ appears as the finite part in the
harmonic series
\be
\sum_{n=1}^\L \frac 1{n}=\ll+\g+O(\L^{-1})\,. 
\ee 
Adding divergences in \rf{diver1}, \rf{diver2}, and \rf{diver3}, we
get the total order $g^2$ divergent contribution to the quadratic
effective action equal to
\be
\fig{c1}{0.17}{0}\Bigl(-\frac {g^2}{16\pi^2}(\ll)^2 -\frac {g^2}{8\pi^2} \ll\Bigr)\,.
\label{ds22}
\ee
Remarkably, the finite part $\g$ canceled in this final result. As we
will see in Section 3.5 below, a similar cancellation of finite
parts of divergent Green's functions occurs on a much larger scale
for general contours, and in fact is essential for the existence of
local counterterms. 

Being proportional to $S_2$, the divergence \rf{ds22} can again be canceled by
a counterterm. This requires the following $O(g)$ correction to \rf{glam}
\be
\frac1{g(\L)
}=\frac 1 g +\frac 1{2\pi}\ll+\frac g{8\pi^2}(\ll)^2 +\frac
g{4\pi^2} \ll\,.\label{glam1}
\ee
An important check of the general structure is provided by the
$(\ll)^2$ term in this expression: its value agrees with the one
following from the 1-loop $\beta$-function as dictated by
renormalizability. The subleading $g\ll$ term corrects 
the $\b$-function
\be
\b(g)=-\frac {g^2}{2\pi}-\frac{g^3}{4\pi^2}\,.\label{betag1}
\ee

Thus we see that to the order that we were able to compute, the
circular contour reparametrization path-integral does indeed define a
renormalizable quantum theory. 
I do not know if a simple proof of renormalizability to all orders 
can be given along the lines of the above argument. 
Perhaps one may use $SL(2,\RR)$ invariance to restrict
the form of the quantum action. At present this has not been done.

In the next section, we will renormalize the path-integral \rf{Z2}
for a general contour, finding a 2-loop $\b$-function agreeing with
\rf{betag1}. This will provide yet another check for our claim of 
renormalizability.
    
\newsubsection{Two-loop renormalization for a general contour}

To compute 2-loop corrections in \rf{Z2}, we must expand the action to
the 4th order in $\b$. It is
convenient (leads to considerable simplifications and manifest
covariance) 
to perform this expansion in terms of the ``normal coordinate''
field $y(\s)$ defined as the invariant length between $c_\mu(\s)$ and
$c_\mu(\s+\b(\s))$
\be
y(\s)=\int_\s^{\s+\b(\s)}\sqrt{\dot{c}^2}\,.
\ee
In terms of $y(\s)$ we have the Taylor expansion
\be
c_\m(\s+\b(\s))=\sum \frac{1}{k!}\,\frac
{d^k c_\m}{ds^k}(\s)\,y^k\,,
\ee
where $s$ is the natural parameter (the $s$ derivatives are denoted
below by a prime). Substituting this into \rf{becomes}, we get 
\be
S   = \Scl[c_\mu(\s)]+S_2+S_3+S_4+\ldots
\ee 
with 
\ba
S_2  \is \dint d\s\,d\s'\, \Ks\, c'_\m y(\s)\, c'_\m y(\s')+\int d\s\,
(\psi  c'')\,y^2,\nonu
\dis S_3 \is \dint \Ks\, c'_\m y(\s)\, c''_\m\, y^2(\s')+\frac1{3}\int
(\psi  c ''')\,y^3\,,\nonu
\dis  S_4 \is \frac 14 \dint \Ks\, c_\m''\, y^2(\s)\,c_\m''\, y^2(\s')\nonu 
&  + \!\!&\frac 13 \dint \Ks\,c_\m'\, y(\s)\,c_\m'''\, y^3(\s') + \frac 1{12}
\int (\psi  c'''')\,y^4\,.
\label{actionparts}
\ea
Here we denoted
\be
\Ks=-\frac{1}{\pi(\s-\s')^2}\,,\qquad \psi_\m=\frac
12\,\frac{\gd\Scl}{\gd c_\m}\,.
\ee
The path-integral \rf{Z2} can now be computed perturbatively
using Feynman rules. The propagator
\be
G(\s,\s')=\int[\cD y]\,y(\s)\,y(\s')\,\exp(-S_2)
\ee
cannot in general be found explicitly. Being the Green's function of
the quadratic action, it satisfies the integral equation
\be
c_\m'(\s)\int d\s' \Ks\, c_\m'(\s')\, G(\s',\s'')\;+\;
(\psi  c'')(\s)\,G(\s,\s'')=\frac 12\, \gd(\s-\s'')\,.
\label{inteq1}
\ee
In the 1-loop approximation the effective action $F=-\log Z$ is given by
\be
F=\frac 1{4\pi\a'}\Scl[c]+\frac 12\,
\fig{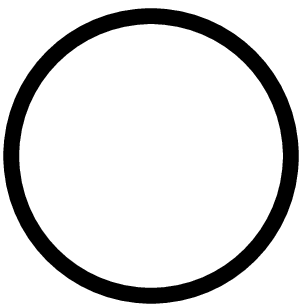}{0.17}{-4.5}\,.
\label{main1}
\ee
As we discussed in Section 2.5, the log-determinant is logarithmically
divergent\footnote{Strictly speaking, there 
is also a linear divergence. But it is contour independent and can be ignored.}
\be 
 \frac 12\,\fig{circle}{0.17}{-4.5}
=\frac {1}{2\pi}\ll\int (\psi  c'')+finite\,.
\label{1l}
\ee
 To cancel this divergence, in Section 3.2 we added a counterterm
renormalizing the contour
\be
\gd c_\m=-\a' \ll\, c_\m''\,.
\ee
Now we would like to move further and consider 2-loop corrections. In
this order the 1-particle irreducible effective action is given by
\NC{\dStwo}{\gd S_2 \hspace{-19.5pt}\fig{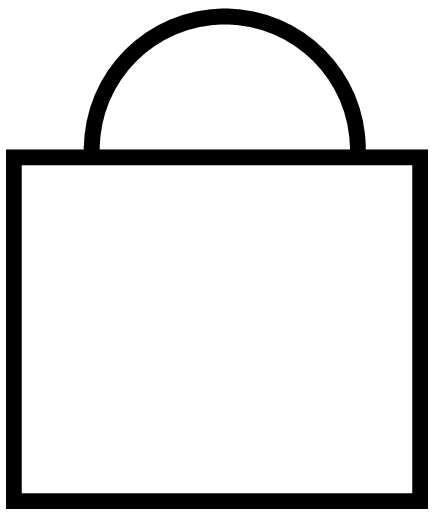}{0.17}{-6.0}}
\NC{\dSfour}{S_4\hspace{-15.5pt}\fig{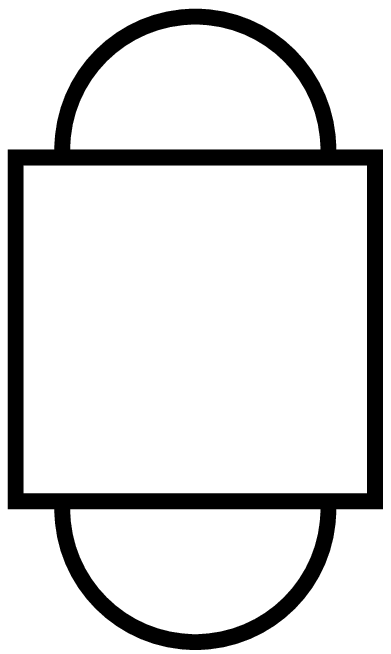}{0.17}{-13}}
\NC{\dSthree}{\makebox[0pt][l]{\fig{box3}{0.17}{-6}}\hspace{3pt}S_3\hspace{14pt}S_3}
\NC{\dSthreea}{\makebox[0pt][l]{\fig{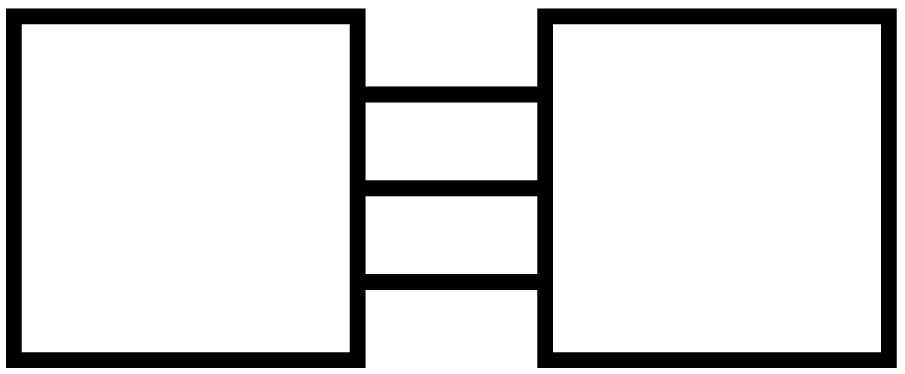}{0.17}{-6}}\hspace{3pt}S_3\hspace{15pt}S_3}
\be 
F_{\rm 1PI}=\frac 1{4\pi\a'} \Scl[c+\gd c] + \frac 12\,\fig{circle}{0.17}{-4.5} +\,\dStwo
+4\pi\a'\Bigl(\ \dSfour- \frac 12\,\dSthreea
\ \Bigr)\,.
\label{F00}
\ee
The change in $S_2$ due to the 1-loop counterterm is
\be
\gd S_2 = 2\dint \Ks\, \gd c_\m'\,y(\s)\,c_\m'y(\s')+\int (\gd
\psi_\m\,c_\m''+\psi_\m\,\gd c_\m'')y^2\,,
\label{ds2}
\ee
where
\ba
\gd\psi_\m(\s)\is-\a'\ll\int d\s'\,\Ks\, c_\m''(\s')\,,\nonu
\gd(c_\m')\is-\a'\ll\Bigl(c_\m'''-c'_\m(c'c''')\Bigr)\,,\nonu
\gd(c_\m'')\is-\a'\ll\Bigl(c_\m''''-2c''_\m c^{\prime\prime 2}-c'_\m(c'c''')'\Bigr)\,.
\label{gdc''}
\ea

We have to analyze divergences in the following diagrams 
\ba
\dStwo \is \fig{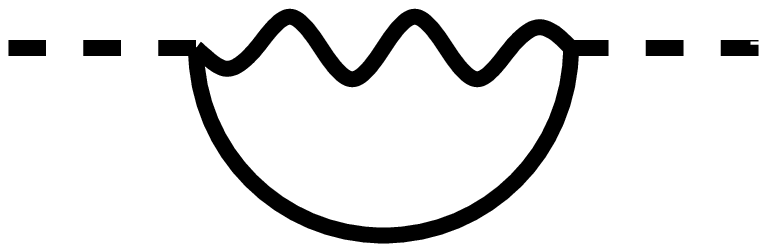}{0.17}{-14.5}+\fig{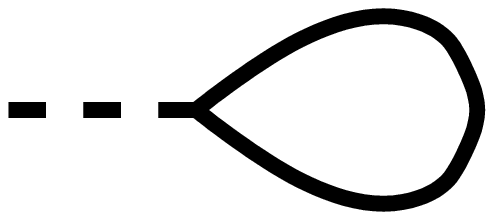}{0.17}{-14.5}
\label{340}\\
\dSfour \is
3\fig{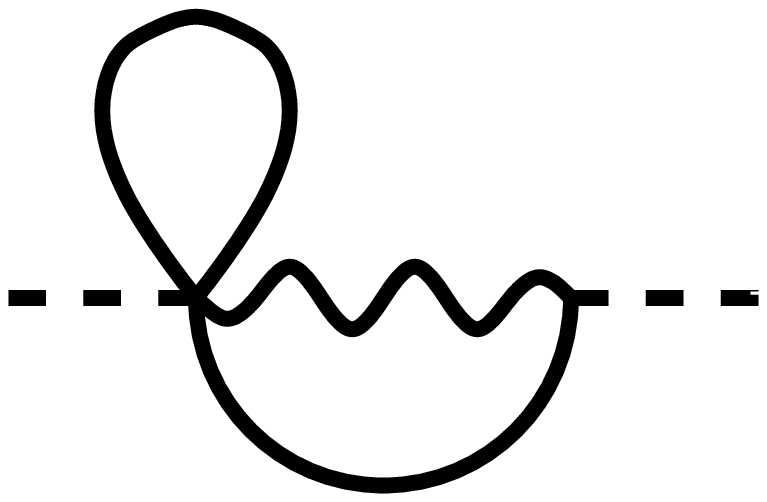}{0.17}{-14.5}+2\fig{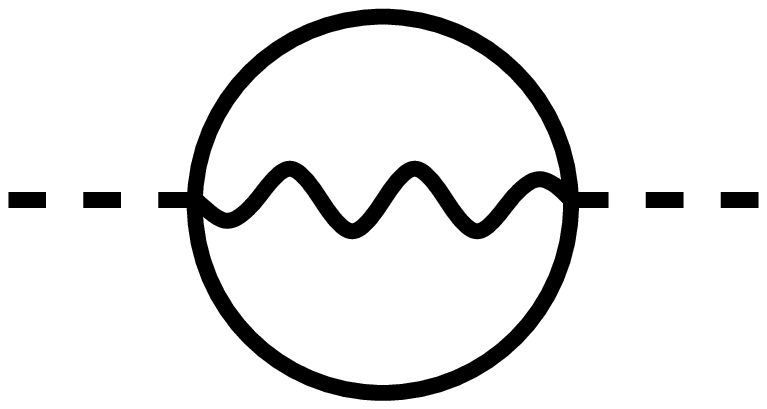}{0.17}{-14.5}+\fig{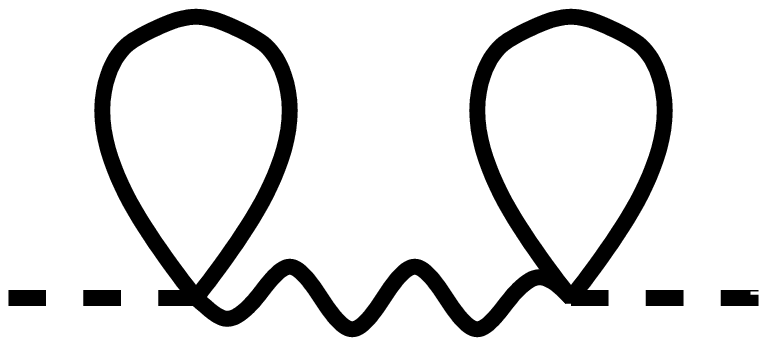}{0.17}{-14.5}+3\fig{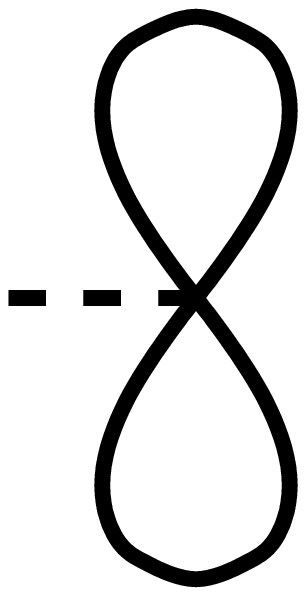}{0.17}{-14.5}
\label{four}\\
\frac 12\;\dSthreea\, \is \fig{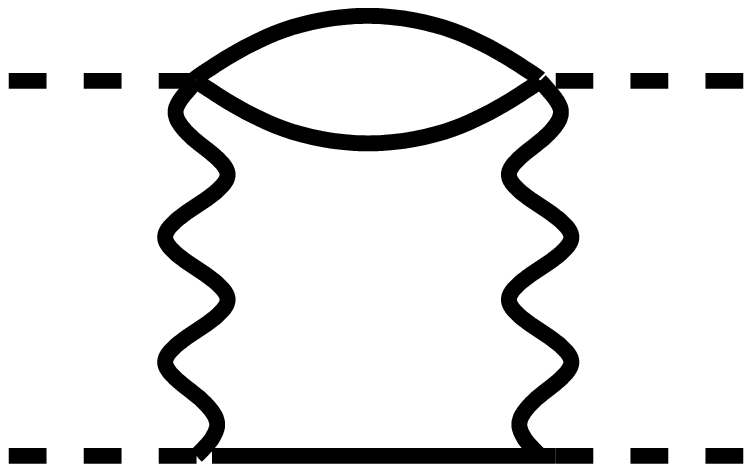}{0.17}{-14.5}+ 2\fig{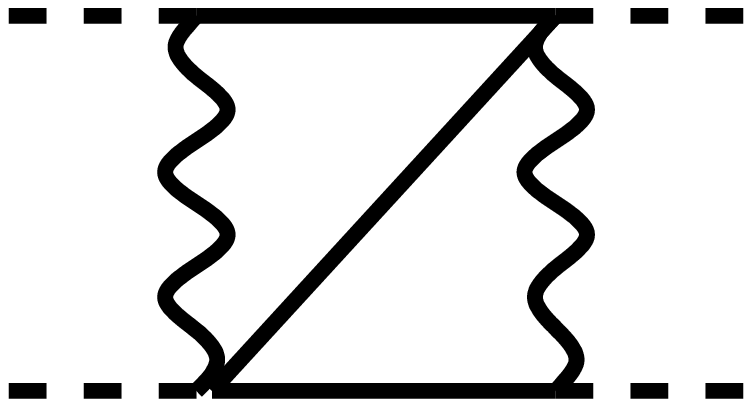}{0.17}{-14.5}+6
\fig{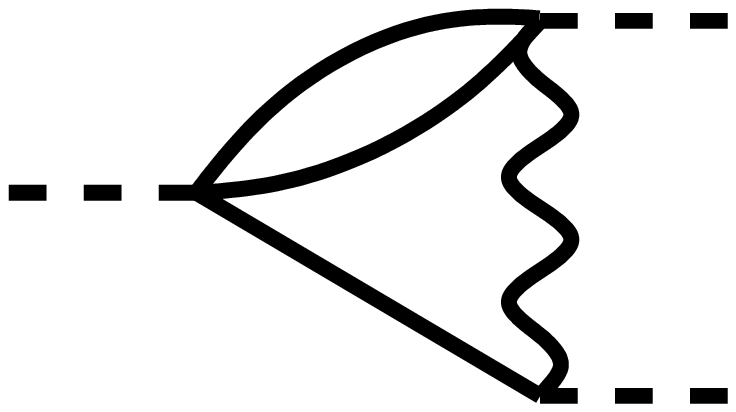}{0.17}{-14.5}+ 3\fig{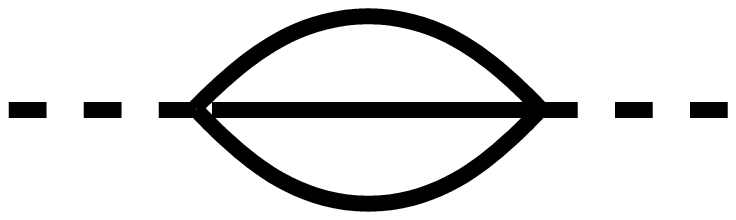}{0.17}{-14.5}
\label{threea}
\ea
Here $ \fig{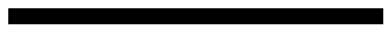}{0.17}{-0.5}=G(\s,\s')$, $
\fig{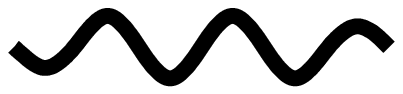}{0.17}{-0.5}=\Ks$, and the dashed lines denote the
$\s$-dependent coefficients present in
\rf{actionparts} and \rf{ds2}. The numerical factors account for
equivalent contractions.

Consider as an example the first diagram in \rf{four}, corresponding to the integral
\be
\fig{S43}{0.17}{-14.5}\,=\,
\frac 13 \dint \Ks\, c'_\m(\s)\,
c_\m'''(\s')\,G(\s,\s')\,G(\s',\s').
\label{3rd0}
\ee
An immediate source of infinities in this expresssion is the presence
of $G(\s',\s')$ in the integrand (the
corresponding part of the diagram is the leaf \fig{leaf}{0.17}{-2}). The regulated Green's
function
at coincident points is logarithmically divergent
\be
G(\s,\s)=\frac{1}{2\pi}\ll+\Gfin(\s).
\label{gr0} 
\ee 
In fact, the propagator has the following asymptotic expansion
\ba
G(\s,p)\is\frac{1}{2|p|}-\frac{(\psi  c'')(\s)}{2p^2}+\ldots\qquad(p\to\infty)\,,\nonu
G(\s,p)&\!\!\stackrel{{\rm def}}{=}\!\!&\int G(\s+\textstyle \frac h2,\s-\frac h2)\,e^{iph}\,dh\,.
\label{G0}
\ea
This formula is not difficult to obtain from
\rf{inteq1}. Eq. \rf{gr0} follows from \rf{G0} with an unknown finite
part which I
denoted $\Gfin(\s)$.

However, the leaf is not the only source of divergence in \rf{3rd0}. A
separate infinity comes from the oyster
\be
\fig{smalloyster}{0.17}{-4}\,\sim\dint\Ks\, G(\s,\s').
\label{oy00}
\ee
This divergence is equal to
\be
\int \frac {dp}{2\pi}\,|p|\,
G(\s,p)=-\frac{1}{2\pi}\ll\,(\psi  c'')(\s)+finite
\label{oy0}
\ee
(ignoring a term linear in $\L$).

Notice that the divergent parts of \rf{gr0} and \rf{oy0} are known
exactly. On the contrary, the finite parts, like $\Gfin(\s)$ in
\rf{gr0}, depend on the data in some complicated nonlocal way and
cannot in general be found explicitly. However, the total divergence
of \rf{3rd0} will arise as the product of \rf{gr0} and \rf{oy0}. So in
addition to the explicitly known $(\log\L)^2$ terms it will contain subleading
$\ll\,$ cross-terms proportional to unknown finite parts. This
presents a potentially serious problem for renormalizability, 
because at the end of the calculation I expect to obtain a local counterterm.

Similar problems with finite parts appear in renormalizing local
field theories, such as $\l\f^4$ theory, in curved space-time. There it
was found \cite{Bunch:1979uk} that divergent contributions containing
finite parts cancel when individual diagrams are added. We will see
below that such a cancellation happens in our case as well.

Careful analysis of divergences in 2-loop diagrams is carried out in
Appendix B along the lines of the above discussion.
The only nontrivial calculations are required for the following three diagrams
\ba
&\dis \frac{1}{4\pi\a'}\fig{S21}{0.17}{-14.5}+3
\fig{S43}{0.17}{-14.5}=\frac{(\ll)^2}{4\pi^2}\int(\psi c'')c^{\prime\prime
2} +\frac{\ll}{\pi}\int(\psi c'')c^{\prime\prime 2}\Gfin\,, \label{theonly1}\\
&\dis2\fig{S42}{0.17}{-14.5}=\Bigl(-\frac{(\ll)^2}{16\pi^2}+\frac{\ll}{8\pi^2}\Bigr)\int
(\psi  c '')c^{\prime\prime2} -\frac{\ll}{4\pi}\int (\psi  c'')c^{\prime\prime2} \Gfin\,
\label{theonly2}
\ea
(it turns out convenient to combine the first two). The diagrams
in \rf{threea} are shown to be convergent. 
All the remaining diagrams do not contain oysters and their
divergences are very easy to write down
using \rf{gr0}. For example
\be
\fig{S41}{0.17}{-14.5}=\frac{(\ll)^2}{16\pi^2}\dint
\Ks\,c''_\m(\s)\,c_\m''(\s')+
\frac{\ll}{4\pi}\dint \Ks\,\Gfin(\s)c''_\m(\s)\,c_\m''(\s')+finite\,.
\ee
We should not also forget about the order $\a'$ term 
\be
\frac 1{4\pi\a'}\dint \Ks\,\gd c_\m(\s)\,\gd c_\m(\s')
\ee
coming from expanding the classical action in \rf{F00}.

I am now going to report the result of these elementary
calculations. The total order $\a'$ divergence in \rf{F00} comes out
to be equal 
\be
\a'\int \frac {(\ll)^2}{4\pi}\Bigl(-(\psi  c'''')+
2 (\psi  c'')c^{\prime\prime 2}\Bigr)+\frac {\ll}{\pi}(\psi  c'')c^{\prime\prime 2} \,.
\label{*00}
\ee
In particular, all the terms proportional to $\Gfin$ indeed cancel
identically in this final result, and the above-mentioned problem happily
resolves itself. Moreover, since the overall divergence is again
proportional to $\psi_\m$, it can be removed by adding an $\a^{\prime 2}$ correction
to the 1-loop counterterm. The total 2-loop counterterm is equal to
\be
\gd c_\m=-\a'\ll\, c_\m''+\frac 12 \a^{\prime 2}(\ll)^2 \Bigl(c_\m''''-
2 c_\m''c^{\prime\prime 2}-c_\m'(c'c''')'\Bigr)-2\a^{\prime 2} \ll\, c_\m''c^{\prime\prime 2}\,.
\label{count00}
\ee
Notice that renormalizability requires that the $(\ll)^2$ coefficient be
expressed via the 1-loop $\b$-function as 
\be
\frac 12\int d\s'\,\frac{\gd \b^\m(\s)}{\gd c_\n(\s')}\,\b^\n(\s')\,.
\ee
To put the coefficient in \rf{count00} in agreement with this
formula, we added a term proportional to $c_\m'$. This new term does
not affect cancellation of divergences, because $(\psi  c')\equiv 0$.  

The $\a^{\prime 2}$ correction to the $\b$-function is given by the $\a^{\prime 2}\ll$ coefficient 
\be
\b^\m(c)=-\Bigl(\a'+2\a^{\prime 2}c^{\prime\prime 2}\Bigr)c_\m''\,.
\ee
This formula is the main result of this section. It is not hard to see
that it agrees with the circular contour 2-loop $\b$-function \rf{betag1}.

This concludes our analysis of the reparametrization path-integral
\rf{Z2}. Obviously we are leaving many
questions open, most notably nonperturbative effects and the precise
relation between the D0-brane and the Wilson loop case (see Section
3.1). I plan to return to these matters in a future publication.
   
\newsection{Reparametrization path-integrals and D$p$-brane dynamics} 

\newsubsection{Nonlocal field theory on the boundary}

The basic conclusion of the previous section was that the D0-brane equations of
motion arise naturally from studying logarithmic divergences in the
reparametrization path-integral \rf{Z2}.

I would now like to consider a generalization of this idea to the case
of D$p$-branes. The D$p$-brane boundary conditions can be written covariantly as 
\be
X^\mu(\s,0)=F^\mu(x^0(\s),\ldots,x^{p}(\s))\qquad(\mu=0\ldots25).
\label{F}
\ee
Here the functions $F^\m(x^a)$ describe the shape of the brane, while the
$p+1$ fields $x^a(\s)$ realize the free boundary
conditions for $X^a$. This is all analogous to \rf{single}, except
that instead of reparametrizations we are now dealing with maps from
$S^1$ into $\RR^{p+1}$. 

The corresponding path-integral generalizing \rf{Z2} is
\ba 
&&\dis \int[\cD x^a(\s)]\,\exp(-\fr{4\pi\a'}\,S\,),\label{part}\\[2mm]
&&\dis S=\dint \Ks\,F^\mu(x(\s))\,F^\mu(x(\s')).
\label{Z}
\ea
From our experience with \rf{Z2}, we expect that the $d=1$ theory of
$p+1$ fields $x^a(\s)$ described by this integral is going to be
renormalizable by adding counterterms changing the shape of the
brane. This is indeed true to the extent that I was able to check it.
Here I am going to present a computation that is technically 
slightly simpler than demonstrating renormalizability to the same
scope as I did it for \rf{Z2}. Namely, we will find a geometric condition on
$F^\mu$ for the cancellation of 2-loop divergences in \rf{part}. This
will give us the D$p$-brane effective action including the first
$\a'$ correction. This is of particular interest, since unlike in the
D0-brane case, the correction will be nontrivial. We will then
compare the result of this calculation with the effective action computed
directly from the $S$-matrix amplitudes, finding complete agreement.
   
\newsubsection{Covariant perturbation theory}

I would like to develop a perturbative expansion of \rf{part}. To begin
with, I choose a configuration of fields $\xbar^a(\s)$ providing a
minimum to the action\footnote{To ensure that such a configuration
exists, I can assume that the brane is asymptotically flat at infinity
and impose the ``long string'' boundary conditions
$x^a(\s)\to\pm\infty$ as $\s\to\pm\infty$. The action of such a long
string coincides with \rf{Z} up to an irrelevant infinite constant.}.
To keep things explicitly covariant, I am going to use 
Riemann normal coordinates with respect to the induced metric
\be
G_{ab}=\d_a F^\mu\, \d_b F^\mu.
\ee
For each $\s$, let $y^a(\s)$ be such coordinates near $\xbar(\s)$. 
I use the normal coordinate field $y^a(\s)$ to represent the field
$x(\s)$, which is supposed to be close to $\xbar(\s)$. In these
coordinates we have the expansion (see e.g. 
\cite{Friedan:1985jm})
\be
F^\mu(x(\s))=F^\mu(\xbar(\s))+\Em_a y^a+\frac12\, \Wm{ab}\,y^ay^b+\frac1{6}\,\LL\,
y^ay^by^c+\frac1{24}\,\MM\, y^ay^by^cy^d+\ldots
\ee
where
\ba
\Em_a = \d_a F^\mu(\xbar(\s)),
&&\LL = \D_{(a}\D_b\,\d_{c)}F^\mu(\xbar(\s)),\nonu
\Wm{ab} = \D_a\,\d_b F^\mu(\xbar(\s)),
&&\MM = \D_{(a}\D_b\D_c\,\d_{d)}F^\mu(\xbar(\s)). 
\label{coeff}
\ea
For future reference note that 
\be
(\Ebar_a\, \Ebar_b)=\gd_{ab},
\ee
since we are in normal coordinates.

Substituting this expansion into \rf{Z}, we get analogously to Section 3.5
\ba
S \is S[\xbar(\s)]+S_2+S_3+S_4+\ldots\,,\nonu
S_2 \is \dint \Ks\, \Em_a y^a(\s)\, \Em_b y^b(\s')+\int
(\Psibar\, \Wbar_{ab})\,y^ay^b,\nonu
S_3 \is \dint \Ks\, \Em_a y^a(\s)\, \Wm{bc}\, y^by^c(\s')+\frac1{3}\int
(\Psibar\,\Lbar_{abc})\,y^ay^by^c,\nonu
 S_4 \is \frac 14 \dint \Ks\, \Wm{ab}\, y^ay^b(\s)\, \Wm{cd}\,
y^cy^d(\s')\nonu
&  + \!\!&\frac 13 \dint \Ks\, \Em_{a}y^a(\s)\, \Lm_{bcd}\,
y^by^cy^d(\s') + \frac 1{12} \int (\Psibar\,\Mbar_{abcd})\,y^ay^by^cy^d,
\ea
where
\be
\Pm(\s)=\int d\s'\,\Ks\,F^\mu(\xbar(\s')),\qquad 
(\Psibar\,\Ebar)\equiv 0\,.
\label{classx}
\ee 
The second order action determines the propagator
$G^{ab}(\s,\s')$. The analogues of \rf{gr0}, \rf{G0} are true
\ba
G^{ab}(\s,p)\is
\frac{\gd^{ab}}{2|p|}-\frac{(\Psibar\,\Wbar_{ab})(\s)}{2p^2}+\ldots\,,\label{gr}\\
G^{ab}(\s,\s)\is \frac{1}{2\pi}\ll\, \gd^{ab} +\Gf{ab}\,.
\label{gr'}
\ea

\newsubsection{Analysis of divergences}

The 1-loop divergence in \rf{part} is easy to find and is given by
 \be 
 \frac 12\,\fig{circle}{0.17}{-4.5}
=\frac {1}{2\pi}\ll\int (\Psibar\,  \Wbar_{aa})\,.
\label{1lx}
\ee
If we want to preserve conformal invariance, this divergence has to
cancel, which requires that the trace of the second fundamental form should vanish to this order 
\be
\Wm{aa}=O(\a').
\label{trace}
\ee
More precisely, this is the most general local condition on the shape
of the brane for which the divergence disappears. 
It is gratifying to see \rf{trace} arise, because this is exactly the minimal surface equation of
motion following from the standard D-brane low energy effective action \cite{Leigh:1989jq}
\be
\int d^{p+1}x\,\sqrt{G}\,.
\ee

Let me now impose Eq. \rf{trace}, so
that there are no order $\a'^0$ divergences in \rf{part}. This does
not mean of course that there will be no $\a'$, $\a'^2$ etc. divergences. In fact, canceling
the divergences to higher and higher order in $\a'$ requires
higher and higher order corrections to the equation of motion. 
Here I am only looking for the first correction. In this order the 1PI
effective action is given by \rf{F00}, except that in our present
situation there is no counterterm $\gd c$, and $\gd S_2$ is also
absent. Thus our job will be simpler than in Section 3.5.

Once again we have to find divergent parts of the diagrams \rf{four} and
 \rf{threea}. This is done similarly to the D0-brane calculations from
 Section 3.5 and Appendix B. Just as before, 
all the diagrams in \rf{threea} will be convergent, because of the
 condition
\be
(\Ebar_a\,\Wbar_{bc})\equiv 0\,.
\ee
Analysis of the remaining diagrams is somewhat simplified 
by the fact that we only need to know the
divergences to order $\a'$. We can use the 1-loop condition
\rf{trace} to show that many terms do not contribute to this order.
To give an example of how this happens, consider the third diagram in
\rf{four}
\be
\fig{S41}{0.17}{-14.5}\,=\,\frac 14 \dint \Ks\, \Wm{ab}(\s)\,G^{ab}(\s,\s)\,
\Wm{cd}(\s')\,G^{cd}(\s',\s').
\ee
Because of \rf{gr'}, all divergences of this diagram will involve terms
like $\Wm{ab}\gd^{\,ab}$, which are $O(\a')$ by \rf{trace}. So this
 diagram is irrelevant.  

The logarithmically divergent parts of the other three diagrams in \rf{four} are
\ba
2\fig{S42}{0.17}{-14.5}\is\Bigl(\frac{(\log\L)^2}{8\pi^2}-\frac{\ll}{4\pi^2}\Bigr)\int(\Psibar\,\Wbar_{ab})\,\R{ab}
\nonumber\\[4mm]
&&-\ \frac{\ll}{2\pi}\int(\Psibar\, \Wbar_{ab})\,\Wm{ac}\Wm{bd}\,\Gf{cd}\ +\
O(\a')
\label{22}
\\[4mm]
3\fig{S43}{0.17}{-14.5}\is -\frac{(\log\L)^2}{6\pi^2}\int(\Psibar\,\Wbar_{ab})\,\R{ab}
-\frac{\ll}{3\pi}\int(\Psibar\,\Wbar_{ab})\,\R{ac}\Gf{bc}\nonumber\\[4mm]
&&+\ \frac{\ll}{6\pi}\int(\Psibar\,\Wbar_{ab})\,\Gf{cd}\Bigl(2\,\Wm{ac}\Wm{bd}+\,\Wm{ab}\Wm{cd}\Bigr)
\ +\ O(\a')
\label{33}\\[4mm]
3\fig{S44}{0.17}{-14.5}\is \frac{(\log\L)^2}{24\pi^2}\int(\Psibar\,\Wbar_{ab})\,\R{ab}
+\frac{\ll}{3\pi}\int(\Psibar\,\Wbar_{ab})\,\R{ac}\Gf{bc}\nonumber\\[4mm]
&&+\ \frac{\ll}{6\pi}\int(\Psibar\,\Wbar_{ab})\,\Gf{cd}\Bigl(\Wm{ac}\Wm{bd}-\,\Wm{ab}\Wm{cd}\Bigr)\
+\ O(\a')\label{44}
\ea
In these formulas $\R{ab}(\s)$ is the Ricci tensor of the induced
metric, written in normal coordinates at the point $\xbar(\s)$. Some
further details about the derivation of these formulas can be found in
Appendix C.

The sum of \rf{22}--\rf{44} is equal
to 
\be
-\,\frac{\ll}{4\pi^2}\int d\s\,(\Psibar\,\Wbar_{ab})\,\R{ab}\ +\ O(\a').
\label{2l}
\ee
In particular, just as we expected, the individual contributions of the unknown finite
parts $\Gf{ab}$ cancel in this sum.

\newsubsection{Corrected equations of motion}

Let me repeat the logic of the above computation. First I noticed that
\rf{part} contains a 1-loop divergence \rf{1lx} of order $\a'^0$. For
this divergence to vanish modulo $\a'$, I had to impose
equation of motion \rf{trace}.
Then I proceeded to compute order $\a'$ divergences coming from the 2-loop
terms. In this computation I actually used \rf{1lx} to show that many
potentially divergent terms are of order $\a'^2$ or higher and can be
ignored.
The result is that the total $\a'$ divergence is
equal to $4\pi\a'$ times \rf{2l}. The divergent part of the 1PI
effective action is
\be
F_{\rm 1PI}^{\rm div}=\rf{1lx}+4\pi\a'\rf{2l}=\frac{\ll}{2\pi}\int d\s\,
\Pm\Bigl(\Wm{aa}-2\a'\,\Wm{ab}\R{ab}\Bigr)\ +\ O(\a'^2).
\ee
Requiring that 
$F^{\rm div}= O(\a'^2)$, I get the corrected equation of motion 
\be 
\Wm{aa}-2\a'\,\Wm{ab}\R{ab}= O(\a'^2).
\ee
It is a simple matter to check that this equation of motion
corresponds to the low energy effective action of the form
\be
 \Seff=T_p\int
d^{p+1}x\,\Bigl(\sqrt{G}+\a'R\sqrt{G}\,\Bigr)+O(\a^{\prime\;2})\,.
\label{seff}
\ee
Thus our computation predicts that there is an order $\a'$ Einstein terms in the
bosonic D$p$-brane effective action, and fixes a relative coefficient
in front of it\footnote{Notice that the Einstein term represents the
only possible correction in this order. Dimensional analysis would also allow for $O(\a')$ terms
containing the second fundamental form $\W^\m_{ab}$ squared. However,
the term $(\W^{\m a}_{a})^2$ is irrelevant on shell, since $\W^{\m a}_a=0$
by the minimal surface equations of motion. The term
$\W^\m_{ab}\W^{\m\,ab}$ is expressible via $R$ by the Gauss-Codazzi
equations.}${}^,$\footnote{After this paper was completed, it was brought to my
attention by Arkady Tseytlin that the Einstein-Hilbert term in the
bosonic D-brane action was previously discovered in 
\cite{Corley:2001hg} by analyzing graviton scattering amplitudes in
the presence of the brane. The relative
coefficient in our formula \rf{seff} agrees with Eq. (3.16) in that paper.}.

It remains to check that this coefficient agrees with 
the direct $S$-matrix analysis, as it should if our interpretation is
correct. 

\newsubsection{ D$p$-brane effective action from the $S$-matrix}

The open string vertex operators 
\be
V^i(q)=\int d\s\, \d_\perp X^i e^{iq_aX^a}\qquad(a=0\ldots p,\
i=p+1\ldots 25)
\ee
describe ripples on the D$p$-brane polarized in the $i$-th transverse
direction and propagating along the brane with momentum $q$. The
4-scattering amplitude for ripples with polarizations
$i,j,k,l$ can be easily computed and is equal to
\ba
&\cA = \gd_{ij}\,\gd_{kl}\, A(s,t,u)+\gd_{ik}\,\gd_{jl}\, A(u,t,s)+\gd_{il}\,\gd_{jk}\, A(t,s,u)\,,\\[2mm]
&A(s,t,u) = B(-1-\a's,1-\a't)+B(-1-\a's,1-\a'u)+B(1-\a'u,1-\a't)\,.\nonumber
\ea 
Expanding in the region of small momenta and keeping the first
$\a'$ correction,
\be
\cA \propto\gd_{ij}\,\gd_{kl}\, (tu-\a' stu)+\ldots+O(\a^{\prime\;2})\,,
\label{ampl}
\ee 
where \ldots\ denotes the terms proportional to the other pairings.

Eq. \rf{ampl} has to be compared with the 4-ripple amplitude following
 from \rf{seff}. To do the comparison, I must expand $\Seff$ to
the second order in $g_{ab}=\d_a\f^i\d_b\f^i$. The actual computation
 is easy to perform using tetrads. The result is 
\ba
\Seff&\!\!\propto\!\!\!\!& \int 1+ \frac12\, g_{aa} + \frac1{8}\, (g_{aa})^2 -\frac1{4}\, g_{ab}\,
g_{ab}\nonu
&&+\,\a'\Bigl(\frac12\, g_{aa}\,\d_b\d_cg_{bc}+\frac12\, \d_a g_{ab}\,\d_c
g_{cb}+\frac1{4}\,g_{ab}\,\d^2g_{ab}-\frac1{4}\,g_{aa}\,\d^2
g_{bb}\Bigr)\!\!+\ldots
\label{aaaa}
\ea
The indices are contracted with $\h_{ab}$ in this formula. 

As the reader may check, the 4-amplitude derived from \rf{aaaa} coincides with (\ref{ampl})
exactly.

\newsection{Conclusion}

In this paper we demonstrated that reparametrization path-integrals
are ubiquitous in string theory problems involving fixed space-time
objects where the strings end, such as D-branes in critical string
theory, or Wilson loops in gauge/string duality. For the first time
we explored quantum properties of these integrals, laying out the 
foundations of any future treatment.

In some cases, such as for the Wilson loop on the boundary of AdS, 
these integrals give a finite contribution to the amplitude. In the
others, such as for the D-branes, the integrals are UV divergent. However, in
the latter case the integrals carry dynamical information related
to the fact that the divergences can be removed by renormalizing the
shape of the brane. This renormalization group flow is governed by the D-brane
low energy effective action.

Many issues related to the above facts were investigated or checked only
partially. Some of the more important open problems are related to the
loop equations (see Section 2.4) and nonperturbative effects (see
Section 3.1). Much further work and insight will be required to
complete the emerging picture.

\vspace{15mm} {\noindent \bf Acknowledgements}

\vspace{4mm}

I would like to express deep gratitude to my advisor Alexander Polyakov 
for the ideas, knowledge, and wisdom he shared with me so generously and 
patiently. He introduced me to the subject, suggested the questions to
study, and helped me interpret the results. 
This paper would have never been written without his constant 
support and attention. 

A part of this work was done while I was attending ``Les Houches 2001''.
I am grateful to the organizers of the school and other participants
for creating a wonderful scientific atmosphere.

I would like to thank Arkady Tseytlin for important comments about the
first version of this paper. I would also like to thank Igor Klebanov and Juan Maldacena 
for the interesting and useful discussions of
gauge/string duality in Les Houches and afterwards.

\setcounter{section}0
\renewcommand{\thesection}{\Alph{section}}

\appsection{ Derivation of \rf{Z2} }

Here we present a careful derivation 
of formula \rf{Z2} from the first principles. To begin with, let us introduce 
a boundary condition for the tangential component of the metric in
\rf{Z0}
\be
\label{y1}
g_{\s\s}(\s,0)=h(\s)\,.
\ee
Thus $Z$ becomes a function of $c^\mu$ and $h$ (although conformal
invariance would mean that the dependence on $h$ is actually not
there.) With these definitions one can think 
of $Z$ as a sort of object obtained by cutting a closed string
world-sheet
into two halves:
\be
\fig{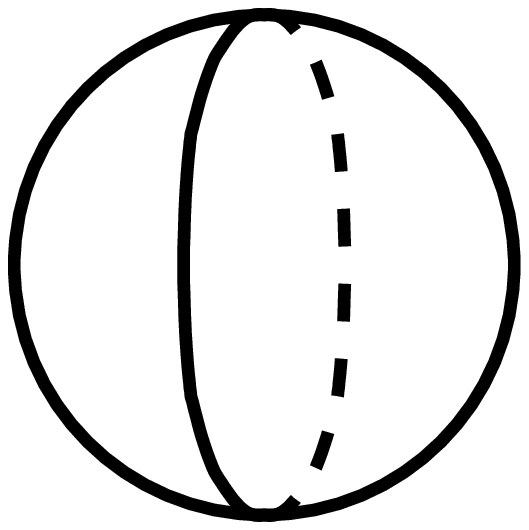}{0.17}{-14}\rightarrow \fig{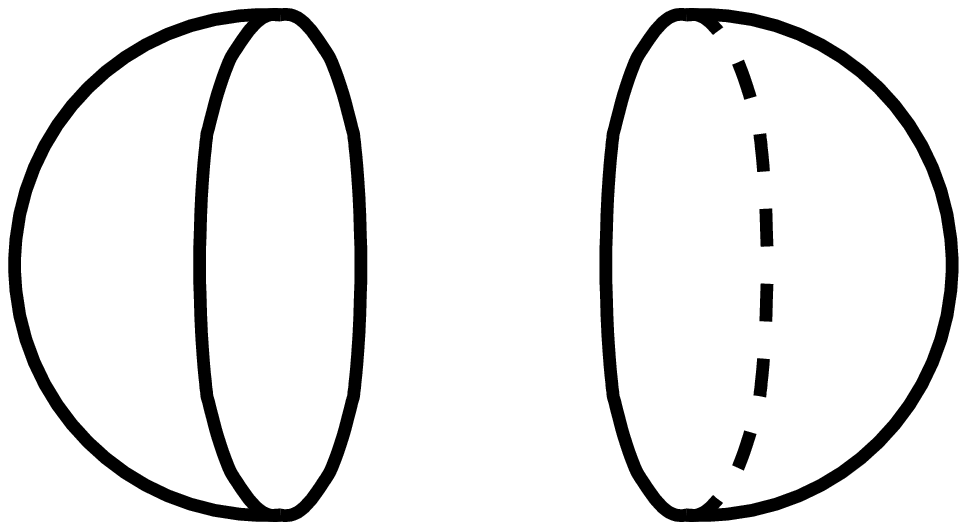}{0.17}{-14}\,.
\ee
In particular, the closed string partition function formally
factorizes
\be
\fig{closed}{0.17}{-14}=\int[\cD c^\mu \cD h]\,(Z[c^\mu,h])^2\,.
\ee

We proceed, as usual, to parametrize the metric $g$ in \rf{Z0} by
\be
g_{ab}=(e^\phi\gd_{ab})^f\,,
\ee
where $f$ is an upper half-plane diffeomorphism bringing $g$ to the
conformal form. Because of \rf{y1}, the pair $(\phi,f)$ is constrained
by the boundary relation
\be
e^{\phi(\b(\s))}[\b'(\s)]^2=h(\s)\,,
\label{y11}
\ee
where $\b(\s)=f^1(\s,0)$. 

Now we change variables 
\be
X^\mu(\xi)=\widetilde{X}^\mu(f(\xi))
\ee
in the path-integral. The new field $\widetilde X$ satisfies the boundary
condition
\be
\widetilde {X}^\mu(\s,0)=c^\mu(\b^{-1}(\s))
\ee
and has the action $\int(\partial \widetilde X)^2$. So the path-integral
over $\cD\widetilde X$ is equal to
\be
\exp(-\Scl[c^\mu( \b^{-1}(\s))])\,,
\ee
where $\Scl$ is given by \rf{scl}. (We ignore conformal anomaly, which
is going to cancel for $D=26$.)
Thus, we get an intermediate result
\be
Z=\int[\cD\phi\cD f]\,\exp(-\Scl[c^\mu( \b^{-1}(\s))])\,.
\label{y21}
\ee

Now we would like to split $\cD f$ into $\cD \b \cD f_0$, where
$f_0\in\Diff_0$ fixes the boundary. This splitting is local and does
not introduce any Jacobian factor. The covariant measure of integration over
$\cD \b$ is 
\be
\|\gd\b\|^2=\int d\s\,e^{3\phi(\s)/2}[\gd\b(\b^{-1}(\s))]^2\,.
\label{y2}
\ee
Finally, let us make change of variables $\b\to\a=\b^{-1}$ in \rf{y21}
and drop the infinite term $\int[\cD \phi\cD f_0]$. This gives us exactly
\rf{Z2}.

The only thing that remains to figure out is the measure of
integration in \rf{Z2}. Using 
\be
\gd\b(\b^{-1}(\s))=-\b'(\b^{-1}(\s))\gd\a(\s)
\ee
and \rf{y11}, it is easy to show that in terms of $\gd\a$ the measure
\rf{y2} takes the form
\be
\|\gd\a\|^2=\int d\s\,h^{3/2}(\s)\,[\gd\a(\a^{-1}(\s))]^2\,.
\label{y3}
\ee
This is the natural covariant measure of integration with fixed
boundary metric $h(\s)$. This assures in particular that $Z$ is invariant with respect
to simultaneous reparametrizations of $c^\mu$ and $h$:
\be
Z[c^\mu,h]=Z[c^\mu(\a(\s)),h(\a(\s))(\a'(\s))^2]\,. 
\ee

In the main text of the paper we chose to suppress the implicit
dependence of $Z$ on $h$ contained in the measure, and detected
conformal anomalies by looking at the logarithmic divergences. An
alternative
way would be to analyze the dependence of the finite part of \rf{Z2}
on $h(\s)$. This dependence comes out to be nonzero and proportional
to
\be
\int d\s\, \frac{\gd\Scl}{\gd
c^\mu(\s)}\frac{\ddot c_\mu}{\dot c^2}\,\log h(\s)
\ee
in the 1-loop approximation. This leads to the same conclusion that
\rf{Z0} is not conformally invariant.

\appsection{ D0-brane divergences}

In this appendix we derive formulas \rf{theonly1} and \rf{theonly2} and explain why the
diagrams in \rf{threea} are convergent. 

The l.h.s. of \rf{theonly1} is equal to
\be
\frac{\ll}{2\pi}\dint\Ks\,G(\s,\s')\,c_\m'(\s)\,c_\m'(c'c''')(\s')+\dint\Ks\,G(\s,\s')\,c_\m'(\s)\,c_\m'''\Gfin(\s')\,.
\label{a1}
\ee
The only divergence of the second term comes from the oyster diagram
\rf{oy00}, \rf{oy0}. This gives us half the second term in the
r.h.s. of \rf{theonly1}.
The first term in \rf{a1} contains both $(\ll)^2$ and $\ll$
divergences. It can in fact be evaluated exactly using the fact that
$G(\s,\s')$ satisfies \rf{inteq1}. Indeed, putting $\s=\s''$ in
\rf{inteq1} gives
\be
c_\m'(\s)\int d\s' \Ks\, c_\m'(\s')\, G(\s',\s)\;+\;
(\psi  c'')(\s)\,G(\s,\s)=\frac 12\, \gd(0)\,.
\label{inteq1'}
\ee
All three terms in this equation are divergent. However, the
divergence in $\gd(0)=\L/2\pi$ is purely linear and can be
discarded. Now we can use \rf{inteq1'} to conclude that in the
regulated theory
\be
c_\m'(\s)\int d\s' \Ks\, c_\m'(\s')\, G(\s',\s)=-\Bigl(\frac{\ll}{2\pi}+\Gfin(\s)\Bigr)
(\psi  c'')(\s)\,.
\label{a2}
\ee
Substituting this into \rf{a1} gives us the first term and the second
half of the second term in the r.h.s. of \rf{theonly1}.

Now let us turn to \rf{theonly2}. This diagram is most conveniently
analyzed in momentum space. 
Since the Green's function $G(\s,p)$ depends on $\s$, the usual Feynman
rules have to be modified. A second momentum $v$ (dual to $\s$) is
associated with every propagator, and $+v/2$ contributes to momentum
conservation in both vertices joined by it. These properties follow
from the representation
\be
G(\s,\s')=\dint \frac {dv\,dp}{(2\pi)^2}\,G(v,p)\,
e^{iv\frac{\s+\s'}2+ip(\s-\s')}.
\ee

The momentum flow in our particular case is
\be
\fig{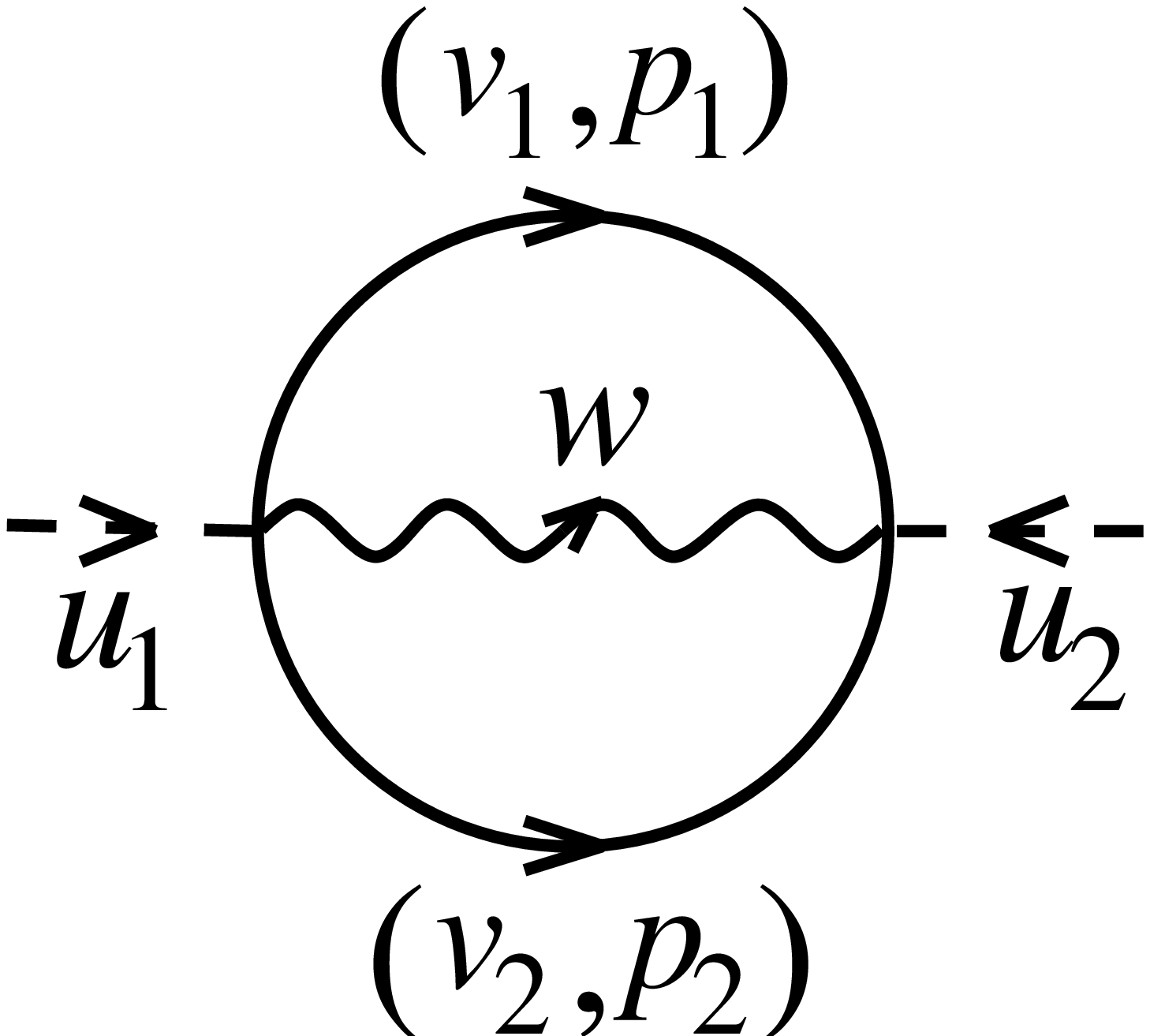}{0.18}{-31}\qquad
\left\{\begin{array}{rcl}
u_1+v_1/2-p_1+v_2/2-p_2-w\is0\,,\\[6pt]
u_2+v_1/2+p_1+v_1/2+p_2+w\is0\,.\\
\end{array}\right.
\ee
Excluding $w$, we get the following expression for the diagram
\ba
\fig{S42}{0.17}{-14.5}\is\frac
14\int\frac{d^2u\,d^2v}{(2\pi)^4}\,2\pi\gd(u_1+u_2+v_1+v_2)\,
c_\m''(u_1)\, c_\m''(u_2)\, J(u_1,u_2,v_1,v_2)\,,\nonumber\\[4pt]
J\is\int\frac{d^2p}{(2\pi)^2}\,
\Bigl|p_1+p_2-u_1-\frac{v_1+v_2}2\Bigr|\,
G(v_1,p_1)\,G(v_2,p_2)\,.
\label{200}
\ea

I regulate the divergent integral $J$ by imposing the cut-off
$|p_i|<\L$. Using \rf{G0} it is easy to see that
\be
 J=\int\frac{d^2p}{(2\pi)^2}\,
|p_1+p_2|\,
G(v_1,p_1)\,G(v_2,p_2)+finite,
\label{J}
\ee
so that the divergent part of $J$ is actually independent of $u_1$ and
$u_2$.

Let us split the region of integration into $|p_2|<|p_1|$ and
$|p_2|>|p_1|$. In the first case
\be
|p_1+p_2|=|p_1|+p_2\,\sign\, p_1.
\ee
It can be shown using \rf{G0} that the second term leads to a convergent
integral in \rf{J}, so that
\be
J_{|p_2|<|p_1|}=\int_{|p_1|<\L}\frac{dp_1}{2\pi}\,|p_1|\,G(v_1,p_1)\int_{|p_2|<|p_1|}\frac{dp_2}{2\pi}\,G(v_2,p_2)+finite.
\label{J1}
\ee
Expansion of the inner integral for $|p_1|\to\infty$ is easy to obtain from \rf{G0}
\be
\int_{|p_2|<|p_1|}\frac{dp_2}{2\pi}\,G(v_2,p_2) =
\gd(v_2)\,\ln |p_1| +
\Gfin(v_2)+\frac{(\psi c'')(v_2)}{2\pi|p_1|} + \ldots
\label{J2}
\ee
The $O(p_1^{-2})$ error term will not contribute to the divergence. By
\rf{G0} and \rf{J2} we know the asymptotics of the integrand in
\rf{J1}. The divergence of $J$ follows from this information in a
straightforward way
\ba
J_{|p_2|<|p_1|}\is-\frac{(\ll)^2}{8\pi^2}\,(\psi c'')(v_1)\,2\pi\gd(v_2)\nonu
&&+\,\frac{\ll}{4\pi^2}\,2\pi\gd(v_1)\,(\psi c'')(v_2) -\frac{\ll}{2\pi}\,(\psi c'')(v_1)\,\Gfin(v_2) +finite,
\label{J3}
\ea
where I also omitted divergent terms proportional to $\L$ and
$\L\log\L$. The $|p_2|>|p_1|$ part of $J$ is obtained from the 
$|p_2|<|p_1|$ part by simply interchanging $v_1\lr v_2$. 

Substituting \rf{J3} into \rf{200}, we get the divergence of the
diagram. For example, the $(\ll)^2$ term comes out to be
\be
-\,\frac{(\ll)^2}{16\pi^2}
\int(\psi c'')c^{\prime\prime 2}
\ee
Analogously all the other terms reduce to single $d\s$
integrals, and we arrive at \rf{theonly2}. 

Finally, let us show that the diagrams in \rf{threea} are convergent. 
For the last diagram it is true simply because
$G^{ab}(\s,p)\sim|p|^{-1}$.
Since $K(p)=|p|$, the other three diagrams in \rf{threea} could in
principle diverge. However, notice that the kernel $\Ks$ appears in
all of them in the combination
\be
A(\s,\s')=c_\m'(\s)\,c_\m''(\s')\Ks\,.
\ee
The coefficient of the leading $p\to\infty$ singularity 
of this expression 
\be
A(\s,p)=(c'c'')(\s)\,|p|+\ldots
\ee
vanishes identically. So in fact
\be
A(\s,p)\sim\sign\, p+\ldots\qquad(p\to\infty)\,,
\ee
from which it follows that the remaining three diagrams in \rf{threea}
are also finite.

\appsection{ D$p$-brane divergences}

D$p$-brane divergences are analyzed analogously to the D0-brane case. 
In this appendix we will indicate
minor differences in the analysis and record Riemannian geometry
identities needed to bring the answers to the form \rf{22}--\rf{44}.

Let us start with the diagram
\ba
\fig{S44}{0.17}{-14.5}\is\frac 1{12}\int d\s\, (\Psibar\,
\Mbar_{abcd})\,G^{ab}(\s,\s)\,G^{cd}(\s,\s)\nonumber\\[4mm]
\is\frac{(\ll)^2}{48\pi^2}\int (\Psibar\, \Mbar_{aabb})+\frac{\ll}{12\pi}\int (\Psibar\,\Mbar_{aabc})\,\Gf{bc}+finite,
\label{4a}
\ea
where I used \rf{gr'} to find the divergent part.

It is not difficult to find by commuting covariant
derivatives that\footnote{The Riemann tensor sign convention is
\ $
(\D_b\D_a-\D_a\D_b)A_d=A_cR^c_{\;d[ab]}.
$
}
\ba
\Mbar^\mu_{aabb}\is\frac 23\, \Wm{ab}\R{ab}+\ldots\nonu
\Mbar^\mu_{aabc}\is\frac 23
\Bigl(\Wm{ad}\Rbar_{abcd}+\Wm{ab}\R{ac}+\Wm{ac}\R{ab}\Bigr)+\ldots
\label{4b}
\ea
where the omitted terms that are either $O(\a')$ by
the use of \rf{trace}, or proportional to $\Ebar^\mu$. 
The latter terms are irrelevant in \rf{4a} because of \rf{classx}.

Now if we express the Riemann tensor by the Gauss-Codazzi equation
\be
\Rbar_{abcd}=\Wn{ac}\Wn{bd}-\Wn{ad}\Wn{bc}\,,
\label{GC}
\ee
we get exactly Eq. \rf{44}.

Let us turn to 
\ba
\fig{S43}{0.17}{-14.5}\is\frac{\ll}{6\pi} \dint \Ks\,G^{ab}(\s,\s') \Em_{a}(\s)\,
\Lm_{bcc}(\s')\nonumber\\[4mm]
&&+\ \frac 13\dint \Ks\,G^{ab}(\s,\s') \Em_{a}(\s)\,
\Lm_{bcd}(\s')\,\Gf{cd}(\s').
\label{30}
\ea
The divergence in the second term is proportional to the oyster
diagram 
\be
\fig{smalloyster}{0.17}{-4}\,=
\int \frac {dp}{2\pi}\,|p|\,
G^{ab}(\s,p)=-\frac{1}{2\pi}\ll\,(\Psibar\,\Wbar_{ab})(\s)+finite
\label{oy}
\ee
and is equal to
\be
-\frac{\ll}{6\pi}\int(\Psibar\, \Wbar_{ab}) \Ebar^\mu_a\Lbar^\mu_{bcd}\,\Gf{cd}.
\label{3a}
\ee
The coefficient $\Lbar^\mu_{bcd}$ can be excluded using the easily derived identity
\be
\Ebar^\mu_a\Lbar^\mu_{bcd}=-\frac13\Bigl(\Wm{ab}\Wm{cd}+\Wm{ac}\Wm{bd}+\Wm{ad}\Wm{bc}\Bigr).
\ee
The same argument applied to the first term in \rf{30} gives me only
the $(\ll)^2$ part of its divergence
\be
-\frac{(\ll)^2}{12\pi}\int(\Psibar\, \Wbar_{ab}) \Ebar^\mu_a\Lbar^\mu_{bcc}.
\label{3b}
\ee
To find the subleading $\ll$ part, a more refined analysis is
required. Commuting derivatives gives
\be
\Lbar^\mu_{bcc}=\frac 23\,\Ebar^\mu_c\R{bc}\ +\ O(\a').
\label{lbar}
\ee
This suggests to invoke the Green's function defining equation
\be
\Em_a(\s)\int d\s' \Ks\, \Em_b(\s')\, G^{bc}(\s',\s'')\;+\;
(\Psibar\,\Wbar_{ab})(\s)\,G^{bc}(\s,\s'')=\frac12\, \gd(\s-\s'')\,\gd_{ac}.
\label{inteq'}
\ee
to evaluate the first integral in \rf{30}. Putting $\s=\s''$ and
ignoring the purely linear divergence coming from $\gd(0)$ gives
\be
\Em_a(\s)\int d\s' \Ks\, \Em_b(\s')\,
G^{bc}(\s',\s)=-\,\frac{\ll}{2\pi}\,(\Psibar\,\Wbar_{ac})(\s)-
(\Psibar\,\Wbar_{ab})\,\Gf{bc}(\s).
\label{3c}
\ee
Using \rf{3c} and \rf{lbar}, I can find the complete logarithmic
divergence of the first term in \rf{30}. It is equal to
\be
-\,\frac{(\ll)^2}{18\pi}\int(\Psibar\,\Wbar_{ab})\,\R{ab}-\frac{\ll}{9\pi}\int(\Psibar\,\Wbar_{ab})\,\R{ac}\,\Gf{bc}\
+\ O(\a').
\label{3d}
\ee
The total divergence of the diagram is equal to the sum of \rf{3a} and
\rf{3d}, and coincides with the answer given in Eq. \rf{33}.

The remaining diagram \rf{22} is analyzed identically to the D0-brane
case. One has to use the relation
\be
\R{ab}=-\Wm{ac}\Wm{bc}\ +\ O(\a')\,,
\ee
which follows from \rf{GC} and \rf{trace}, to transform the answer to the
form given in \rf{22}. I omit the details.

\vspace{20pt}

\renewcommand{\Large}{\normalsize}
\bibliographystyle{rytchkov1}
\bibliography{reparam}
\end{document}